\documentclass[twocolumn,nofootinbib,floatfix,amsmath,amssymb,showpacs,superscriptaddress]{revtex4}

\usepackage{graphicx}
\usepackage{dcolumn}
\usepackage{bm}
\usepackage{color}

\newcommand{\dqex}{\rm ex}

\begin{document}

\title{Hadron-hadron interaction from SU(2) lattice QCD}

\author{Toru T. Takahashi}
\affiliation{Yukawa Institute for Theoretical Physics, Kyoto University,
Sakyo, Kyoto 606-8502, Japan}
\affiliation{Gunma National College of Technology, Maebashi, Gunma
371-8530, Japan}
\author{Yoshiko Kanada-En'yo}
\affiliation{Yukawa Institute for Theoretical Physics, Kyoto University,
Sakyo, Kyoto 606-8502, Japan}
\affiliation{Department of Physics, Kyoto University, 
Sakyo, Kyoto 606-8502, Japan}

\date{\today}

\begin{abstract}
We evaluate interhadron interactions in two-color lattice QCD from
Bethe-Salpeter amplitudes on the Euclidean lattice.
The simulations are performed in quenched SU(2) QCD
with the plaquette gauge action at $\beta = 2.45$ and the Wilson quark action.
We concentrate on S-wave scattering states of two scalar diquarks.
Evaluating different flavor combinations with various quark masses,
we try to find out the ingredients in hadronic interactions.
Between two scalar diquarks
($u C\gamma_5 d$, the lightest baryon in SU(2) system),
we observe repulsion in short-range region,
even though present quark masses are not very light.
We define and evaluate the ``quark-exchange part'' in the interaction,
which is induced by adding quark-exchange diagrams,
or equivalently, by introducing Pauli blocking among some of quarks.
The repulsive force in short-distance region
arises only from the ``quark-exchange part'', and
disappears when quark-exchange diagrams are omitted.
We find that the strength of
repulsion grows in light quark-mass regime and its quark-mass dependence is
similar to or slightly stronger than that of the color-magnetic 
interaction by one-gluon-exchange (OGE) processes. 
It is qualitatively consistent 
with the constituent-quark model picture that 
a color-magnetic interaction among quarks is the origin of repulsion. 
We also find a universal long-range attractive force, 
which enters in any flavor channels of 
two scalar diquarks and whose interaction range 
and strength are quark-mass independent. 
The weak quark-mass dependence of interaction ranges in each component
implies that meson-exchange contributions 
are small and subdominant, and the other contributions,
{\it ex.} flavor exchange processes, 
color-Coulomb or color-magnetic interactions, 
are considered to be predominant, in the quark-mass range we evaluated.
\end{abstract}
\maketitle

\section{Introduction}

Hadron-hadron interactions play important roles in nuclear and hadron physics.
Among them baryon-baryon interaction is one of the most important
issues to be clarified in reliable manners: Baryons are the 
main building blocks of our world and 
their interactions are essential for the structures and dynamics of nuclei.
After the successful proposal by Yukawa in 1935~\cite{Yukawa:1935xg},
the long- and the intermediate-range interactions have been
described by one-boson-exchange potentials (OBEP)~\cite{PTPS39:1967,RMP39:1967,Machleidt:1987hj}.
The key players for the hadronic interactions 
in these regions are considered as mesonic modes.
On the other hand, the short-range part of the interactions
has not been well clarified so far. Especially, the origin of 
the large repulsive core in nucleon-nucleon (NN) channel~\cite{Jastrow:1951},
which accounts for the high-density nuclear phenomena,
has been one of the longstanding problems 
in hadron physics~\cite{Otsuki:1965yk,Neudachin:1977vt,PTPS39:1967,Liberman:1977qs,Harvey:1980rva,Oka:1980ax,Oka:1981ri,Oka:1981rj,Faessler:1982ik,Oka:1984,Oka:2000wj}.
The degrees of freedom of quarks and gluons
would be essential in such short-distance region,
and hence they should be clarified directly in terms of 
the fundamental theory, Quantum ChromoDynamics (QCD).

It is now common knowledge that
all the hadronic systems are governed by QCD,
and it is long after QCD was established as the fundamental theory.
Nevertheless, 
QCD-based description of hadronic systems has not yet been successful
due to the strong coupling nature of the low-energy QCD,
whose dynamics is far from the perturbative regime 
and essentially nonperturbative.
Among potentially workable strategies for nonperturbative analyses of QCD,
lattice QCD is a unique reliable method
for low-energy QCD.
It can now reproduce empirical hadronic masses with a very good 
accuracy~\cite{Aoki:2008sm},
and its outcome can be directly compared with experiments.
There have been several attempts which aim at
clarification of interhadron interactions 
by means of lattice QCD~\cite{Mihaly:1996ue,Stewart:1998hk,Michael:1999nq,Takahashi:2006er,Doi:2006kx}.
Recently, a nucleon-nucleon (NN) interaction was evaluated
from NN Bethe-Salpeter (BS) amplitudes on the lattice~\cite{Ishii:2006ec}, 
where the short-range repulsion and the attraction
in the intermediate region were observed.
While the information of scattering phase shifts are properly encoded
in asymptotic BS amplitudes,
``potentials'' directly constructed from BS amplitudes
have setup- (operator- or energy-) dependences.
Nevertheless, they are considered useful 
to gain qualitative understanding of hadronic interactions.

In this article, we employ SU(2) lattice QCD,
and investigate hadronic interactions aiming at clarifying
the essential structure of them.
First, 
the scalar diquark ($u C\gamma_5 d$, the lightest baryon in SU(2) system)
is an isospin-zero scalar hadronic state in SU(2) QCD
and some of possible one-meson-exchange channels in diquark
interactions are restricted, and hence analyses could be simpler.
Diquarks in SU(2) QCD have a direct connection to mesons,
and two-diquark correlators are identical with
two-meson correlators if we neglect possible quark-disconnected diagrams.
Second, while SU(2) and SU(3) QCDs have different natures
in some situations~\cite{Kogut:1999iv,Kogut:2001na},
they often show similar aspects~\cite{Bali:1994de},
and hence SU(2) QCD can be a testbed for understanding hadron interactions.
On the lattice, it is possible to
switch on/off Pauli-blocking effects among quarks
controlling flavor contents or equivalently quark-exchange diagrams,
which enables us to classify hadronic interactions in terms of their origins.

The paper is organized as follows.
In Sec.~\ref{SecFormulation},
we give a brief explanation of our strategy.
Lattice QCD results are shown in Sec.\ref{SecLattice},
and Sec.\ref{SecDiscussions} is devoted to discussions.
In Sec.\ref{SecSome}, we show the results of additional trial analyses.
The summary and conclusions are given in Sec.~\ref{SecSummary}.

\section{Formulation}
\label{SecFormulation}

We follow the strategy proposed by CP-PACS group~\cite{Aoki:2005uf},
where asymptotic Bethe-Salpeter wavefunctions on the Euclidean lattice
are adopted to evaluate pion scattering length.
This method has been further employed 
for various hadron-hadron channels~\cite{Sasaki:2008sv,Nemura:2008sp}.

We measure a two-diquark correlator $W_{ij,kl,\Gamma}({\bf R},T)$ 
at Euclidean-time $t$, 
which is defined as
\begin{eqnarray}
&&W_{ij,kl,\Gamma}({\bf R},t)
\equiv\\
&&\sum_{\bf x}
\langle
D_{ij,\Gamma}({\bf x},t) D_{kl,\Gamma}({\bf x}+{\bf R},t)
D_{ij,\Gamma}^\dagger (0,0)D_{kl,\Gamma}^\dagger(0,0)
\rangle.
\end{eqnarray}
Here, ${\bf R}$ represents the relative coordinate of two scattering hadrons,
and $D_{ij,\Gamma}({\bf x},t)$ and $D_{ij,\Gamma}^\dagger({\bf x},t)$
are interpolating fields for diquark states
whose flavors are $i$ and $j$.
$D_{ij,\Gamma}({\bf x},t)$ is defined with two quark fields,
\begin{equation}
D_{ij,\Gamma}({\bf x},t)
\equiv
\varepsilon^{ab} q^a_i({\bf x},t) \Gamma q^b_j({\bf x},t),
\end{equation}
with $\varepsilon^{ab}$ being the $2\times 2$ anti-symmetric tensor.
Spinor indices of the two quark fields in $D$ are contracted with $\Gamma$.
Possible $\Gamma$'s are $C$, $C\gamma_5$, $C\gamma_\mu$ and $C\gamma_\mu\gamma_5$,
which respectively 
correspond to pseudo-scalar, scalar, axialvector, and vector diquarks.
We employ wall-type operators for sources, while
we use point-type operators for sinks.
After Euclidean-time evolution,
a system is finally dominated by the ground-state, and
$W({\bf R},t)$ loses $t$-dependence besides an overall constant
and an exponential damping factor.
The BS wavefunction $W({\bf R})$ is then obtained as
$W({\bf R}) \equiv \lim_{t\rightarrow\infty}W({\bf R},t)$.

With such $W({\bf R})$,
we can define an ${\bf R}$-dependent function $V({\bf R})$
assuming a nonrelativistic Schr{\" o}dinger-type equation,
\begin{equation}
\left(
\frac{{\bf p}^2}{2\mu}+V({\bf R})
-E
\right)
W({\bf R})
=0.
\end{equation}
Thus extracted scattering-energy dependent function $V({\bf R})$
is an equivalent potential which reproduces the scattering phase shift
at the target energy.
Though the information of phase shift
is properly encoded in such equivalent potentials $V({\bf R})$,
they could be
energy- and operator-dependent.
In this sense, equivalent potentials $V({\bf R})$
contain no more definite information than phase shifts in asymptotic region.
Nevertheless, 
equivalent potentials are expected to be useful 
to gain understanding  of hadronic interactions,
at least qualitatively.
We extract equivalent potentials,
which we simply call as ``potential'' in this paper,
and evaluate hadronic interactions in SU(2) QCD.

We concentrate on S-wave scattering states of
two scalar diquarks ($\Gamma = C\gamma_5$),
and we project wavefunctions $W({\bf R})$ 
onto $A_1^+$-wavefunction $W(R)\equiv W(|{\bf R}|)$,
which has overlap with $l=0$ states,
by summing over ${\bf R}$ in terms of corresponding discrete rotations.
We here neglect the contributions from $l\geq 4$ scattering states,
since such contributions can be dropped by taking large Euclidean time $t$.

All the simulations are performed in SU(2) quenched QCD
with the standard plaquette gauge action and the Wilson quark action.
The lattice size is $24^3\times 64$ at $\beta = 2.45$,
whose lattice spacing is about 0.1 fm
if we assume $\sqrt{\sigma}$ is 440 MeV~\cite{Fingberg:1992ju,Stack:1994wm}.
We employ four different Hopping parameters 
$\kappa$ = 0.1350, 0.1400, 0.1450, 0.1500 for quarks.
Diquarks at these $\kappa$'s would be well described by
conventional quark models, and Nambu-Goldstone-boson nature
does not emerge. (See Secs.~\ref{SecLatticeHadron} and \ref{SecRemoval}.)

\section{Lattice QCD results}
\label{SecLattice}
\subsection{Hadron masses}
\label{SecLatticeHadron}

We show the lowest-state diquark masses 
in each channel in Table~\ref{hadronicmasses}.
The masses are extracted in a standard manner.
We fit hadronic correlators $C(t)\equiv \langle H(t) H^\dagger(t_{\rm src})\rangle$
by a single-exponential function, 
$C(t)=C\exp(-mt)$.
We do not need ``quark-annihilation'' diagrams
in the computation of diquark correlators.
The scalar, vector, pseudoscalar, axialvector
($\Gamma = C\gamma_5,\  C\gamma_\mu\gamma_5,\  C,\  C\gamma_\mu$) channels
are investigated.

\begin{figure}[h]
\begin{center}
\includegraphics[scale=0.26]{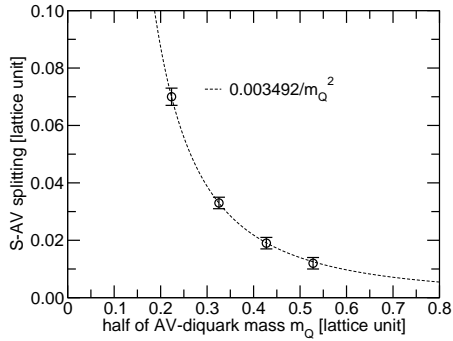}
\end{center}
\caption{
Scalar-axialvector mass splitting is plotted 
as a function of axialvector-diquark mass.
The dotted line is a fit function $Cm_{\rm Q}^{-2}$.
\label{Figpirho}}
\end{figure}

Fig.~\ref{Figpirho} shows the mass difference $\Delta m$ between scalar and axialvector diquarks.
The dashed line denotes the fit function, $\Delta m=C m_{\rm Q}^{-2}$,
where $m_{\rm Q} \equiv \frac12 m_{\rm PV}$ is 
the half of the axialvector-diquark (vector-meson) mass.
The mass splitting $\Delta m$ is clearly proportional to $m_{\rm Q}^{-2}$, which 
supports the color-magnetic interaction as the origin of the mass splitting
and indicates the validity of nonrelativistic-quark-model description
of scalar and axialvector diquarks, at least in the quark-mass range 
we consider.
In the picture of constituent quark model,
if spatial wave functions of constituent quarks in diquarks
are affected by quark-mass variation,
the $m_{\rm Q}^{-2}$ dependence of the mass splitting can be modified
from the $m_{\rm Q}^{-2}$ scaling, which originates 
from the factor $m_{\rm Q}^{-2}$
of the strength of the color-magnetic interaction.
The almost perfect $m_{\rm Q}^{-2}$-behavior 
of the mass splitting
then implies that 
the spatial wave functions in these two diquarks are similar and less dependent
on quark mass.

\begin{table}[h]
\begin{tabular}{llllll}
\hline
$\kappa$ & Scalar & Axialvector & Pseudoscalar & Vector & $\Delta m$\\ \hline\hline
0.1350            & 1.044(2) & 1.056(2) & 1.285( 2) & 1.286( 3) & 0.012(2)
\\
0.1400            & 0.836(2) & 0.855(2) & 1.102( 2) & 1.102( 4) & 0.019(2)
\\
0.1450            & 0.618(2) & 0.651(2) & 0.919( 4) & 0.918( 5) & 0.033(2)
\\
0.1500            & 0.377(3) & 0.447(2) & 0.757( 7) & 0.728( 4) & 0.070(3)
\\ \hline\hline
0.1150            & 1.020(2) & 1.026(2) & 1.397( 9) & 1.352( 8) & 0.006(2)
\\
0.1250            & 0.494(2) & 0.504(1) & 1.053(25) & 0.889(19) & 0.010(2)
\\ \hline\hline
\end{tabular}
\caption{\label{hadronicmasses}
All the hadronic masses are listed.
The masses at $\kappa$=0.1150 and 0.1250 are obtained without
high-energy gluons. (See Sec.~\ref{SecRemoval}.)
$\Delta m$ represents the scalar-axialvector diquark mass splitting.
}
\end{table}

\subsection{Wavefunctions}

We hereby consider two-scalar-diquark wavefunctions with
two different flavor combinations.
One combination is $(i,j,k,l)=(1,2,1,2)$,
where only two independent flavors exist in four quarks.
The other is $(i,j,k,l)=(1,2,3,4)$,
where all the quarks have different flavors.
We adopt the same hopping parameters (quark masses) 
for all the quarks in both cases,
and therefore all the scalar diquarks degenerate in mass.
We note that the diagrams for the $(i,j,k,l)=(1,2,1,2)$ channel
consist of direct and quark-exchange diagrams,
and the direct diagrams in the $(i,j,k,l)=(1,2,1,2)$ channel
are identical with those needed for the $(i,j,k,l)=(1,2,3,4)$ channel.
``Quark-exchange diagrams'' in quark-propagator contraction
are included only in the $(i,j,k,l)=(1,2,1,2)$ case.
The existence of quark-exchange diagrams
would be essential for the short-range interactions,
since any Pauli-blocking effects among quarks are not included
without exchange diagrams.
We expect that the origin of short-range hadronic interactions
can be accessed by comparing these two flavor combinations.

\begin{figure}[h]
\begin{center}
\includegraphics[scale=0.26]{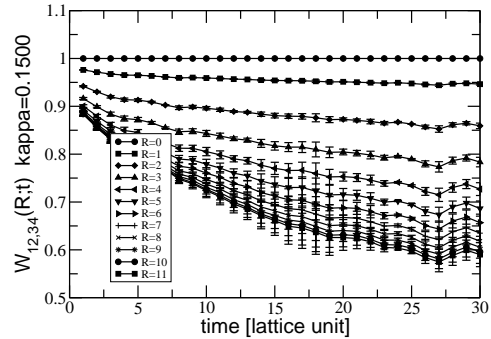}
\includegraphics[scale=0.26]{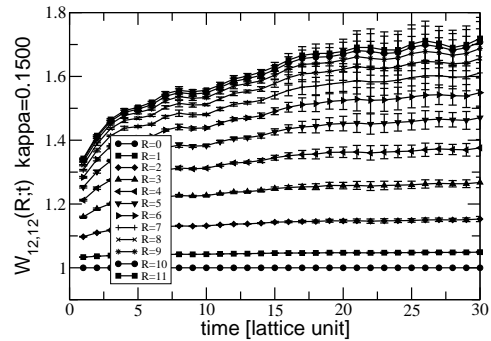}
\end{center}
\caption{
\label{Figt-dependence0}
The wavefunctions $W(R,t)$ 
at $\kappa$=0.1500 are plotted 
as a function of a source-sink separation $t$.
}
\end{figure}

\begin{figure}[h]
\begin{center}
\includegraphics[scale=0.28]{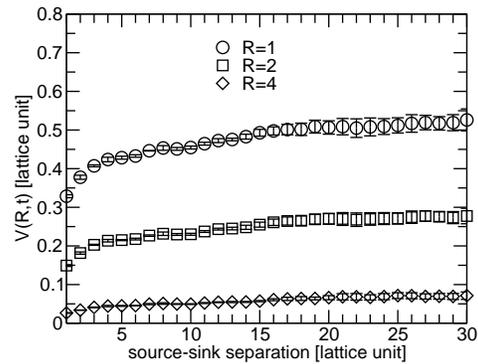}
\end{center}
\caption{
\label{Figt-dependence}
The potentials $V(R,t)$ 
at $\kappa$=0.1500 and $R$=1, 2, 4 are plotted 
as a function of a source-sink separation $t$.
}
\end{figure}

We refer to the determination of wavefunctions $W(R)$.
Law wavefunctions $W(R,t)$ depend on the source-sink separation $t$
due to possible excited-state contaminations.
We then have to take enough large $t$
to ensure the absence of such contaminations.
For this aim, we extract $t$-dependent wavefunctions $W(R,t)$
and determine $t$-window where excited-state contaminations are negligible.
Fig.~\ref{Figt-dependence0} shows
$t$-dependent wavefunctions $W(R,t)$
at $\kappa$=0.1500
as functions of source-sink separation $t$.
The upper and lower panels correspond to $(i,j,k,l)=(1,2,3,4)$ 
and $(1,2,1,2)$ cases, respectively.
They show plateaus at $t \geq 25$, 
and we then extract ``wavefunctions'' $W(R)$
by fitting the data as $W(R,t)=W(R)$ at $t \geq 25$.
For further confirmation,
we look at $t$-dependent potentials $V(R,t)$ extracted
directly from $W(R,t)$.
In Fig.~\ref{Figt-dependence},
$t$-dependent potentials $V(R,t)$ ($R$=1,2,4) at $\kappa$=0.1500
are shown.
One can find that they consistently show plateaus at $t \geq 25$.

\begin{figure}[h]
\begin{center}
\includegraphics[scale=0.28]{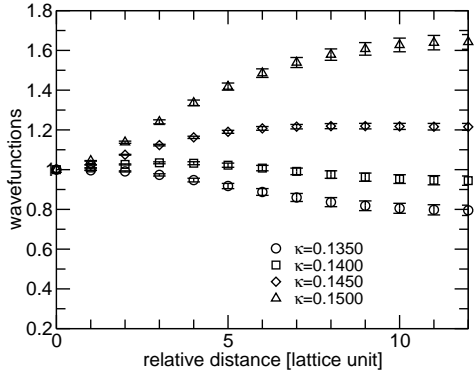}
\end{center}
\caption{
\label{Figwavefunctions}
The wavefunctions $W(R)$ 
for the flavor combination $(i,j,k,l)=(1,2,1,2)$
are plotted 
as functions of relative distance $R$
for $\kappa$=0.1350, 0.1400, 0.1450 and 0.1500.
The corresponding hadronic masses can be found in Table~\ref{hadronicmasses}.
}
\end{figure}
In Fig.~\ref{Figwavefunctions}, as an example,
we show the wavefunctions $W(R,25)$ as functions of the relative distance $R$
for four different $\kappa$'s.The flavor combinations are all
$(i,j,k,l)=(1,2,1,2)$.

\subsection{Potentials : $(i,j,k,l)=(1,2,3,4)$ case}

\begin{figure}[h]
\begin{center}
\includegraphics[scale=0.3]{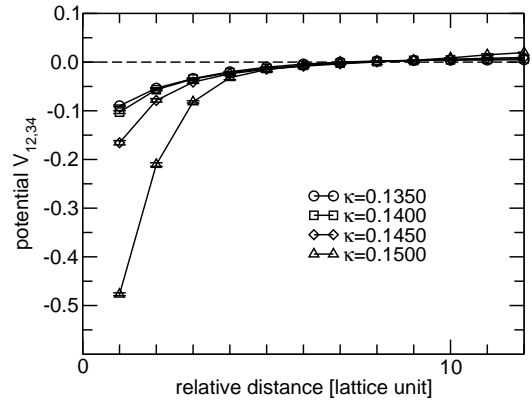}
\end{center}
\caption{
\label{Figpotentials_diag1}
Potentials $V_{12,34}(R)$ computed 
with the flavor combination, $(i,j,k,l)=(1,2,3,4)$,
are plotted as functions of relative distance $R$.
}
\end{figure}

We next proceed with the potentials extracted from the wavefunctions.
Fig.~\ref{Figpotentials_diag1} shows
the reconstructed potentials plotted as functions of $R$,
where the flavor combination is set to $(i,j,k,l)=(1,2,3,4)$.
In this case, we include no quark-exchange diagram
and hence no Pauli-blocking effect among quarks is activated.

At a glance, one can find that 
the interaction in this channel is always attractive,
and that the strength of this attraction depends on quark masses.
The $R$-dependence is monotonous at all the $\kappa$'s.
In the large $R$ region ($R\geq 4$),
the potential has smaller $m_q$ dependence,
while it has strong $m_q$ dependence
in the short-distance region ($R\leq 4$).
The potential $V(R)$ rapidly reduces 
with decreasing $\kappa$ (increasing $m_q$),
and finally
the potentials at $\kappa$=0.1350 and 0.1400
coincide with each other.
The potential exhibits saturation at heavy quark-mass region,
which implies a long-range {\it quark-mass independent} attraction.
Although quark masses are also responsible for hadron size itself
and would indirectly affect interhadron potentials,
these effects seem small in the present quark-mass range.

\subsection{Potentials : $(i,j,k,l)=(1,2,1,2)$ case}

\begin{figure}[h]
\begin{center}
\includegraphics[scale=0.3]{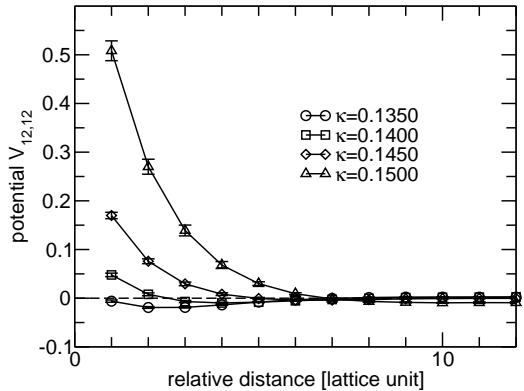}
\end{center}
\caption{
Potentials $V_{12,12}(R)$ computed 
with the flavor combination, $(i,j,k,l)=(1,2,1,2)$,
are plotted as functions of relative distance $R$.
\label{Figpotentials_diag12}}
\end{figure}

The interhadron potentials with the flavor combination
$(i,j,k,l)=(1,2,1,2)$ can be seen in Fig.~\ref{Figpotentials_diag12}.
The difference from the previous $(i,j,k,l)=(1,2,3,4)$ case
is that quark-exchange diagrams are now included,
which gives rise to Pauli-blocking effect among quarks.

We readily find that strong repulsions in the short-distance region
appear in this case.
Compared with the $(i,j,k,l)=(1,2,3,4)$ case,
the quark-mass dependence is not monotonous:
They are smooth functions of $R$ at large $\kappa$'s,
while the potential at $\kappa$=0.1350 
has a pocket at the intermediate distance.
As the quark mass decreases, the short-range repulsive part
rapidly grows up and the intermediate attractive pocket disappears.
Such qualitative change in shape
suggests that the potentials in $(i,j,k,l)=(1,2,1,2)$ case consist of 
two or more parts;
attractive part and (probably) 
repulsive part whose strengths and quark-mass dependences
are different from each other.

\subsection{Potentials : Quark-exchange part}

\begin{figure}[h]
\begin{center}
\includegraphics[scale=0.3]{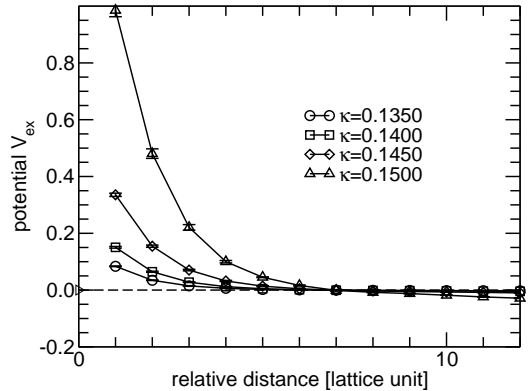}
\end{center}
\caption{
\label{Figpotentials_diff}
``Quark-exchange parts'' of potentials,
which are defined as 
$V_{\dqex}(R)\equiv V_{12,12}(R)-V_{12,34}(R)$,
are plotted as functions of relative distance $R$.
}
\end{figure}

In the following, we define a ``quark-exchange part'' $V_{\dqex}(R,m_q)$ in
$V_{12,12}(R,m_q)$ as
\begin{eqnarray}
V_{\dqex}(R,m_q)\equiv 
V_{12,12}(R,m_q) - V_{12,34}(R,m_q),
\end{eqnarray}
which is nothing but the difference
between $V_{12,12}(R,m_q)$ and $V_{12,34}(R,m_q)$.
Taking into account that quark-exchange diagrams are included
only in the case of $V_{12,12}(R,m_q)$
and that the diagrams needed for $V_{12,34}(R,m_q)$ are identical with
the direct diagrams in $V_{12,12}(R,m_q)$,
it gives
a measure of ``quark-exchange effect'' or Pauli-blocking effect among quarks
in the potential $V_{12,12}(R,m_q)$.
In practice, it is equivalent to assuming
\begin{eqnarray}
&&V_{12,34}(R,m_q)=V_{\rm dir}(R,m_q), \\
&&V_{12,12}(R,m_q)=V_{\rm dir}(R,m_q)+V_{\dqex}(R,m_q),
\end{eqnarray}
for $V_{12,34}(R,m_q)$ and $V_{12,12}(R,m_q)$,
with a ``direct part'' $V_{\rm dir}(R,m_q)$
measured only with direct diagrams
and a ``quark-exchange part'' $V_{\dqex}(R,m_q)$
induced by adding exchange diagrams
or equivalently by introducing Pauli blocking among quarks.
Such decomposition of a diquark-diquark potential
is conceptually similar to the usual decomposition procedure
in RGM (Resonating Group Method) 
calculations in quark cluster models~\cite{Koike:1986mm,Oka:1981ri,Oka:1981rj}.
In case multiply repeated quark-exchange processes
can be neglected, like the Born approximation,
these two components, $V_{\rm dir}$ and $V_{\dqex}$,
result in so-called direct and exchange potentials, respectively.
Thus extracted potential
$V_{\dqex}(R,m_q)\equiv V_{12,12}(R,m_q)-V_{12,34}(R,m_q)$
shows a monotonous behavior as is seen in Fig.~\ref{Figpotentials_diff},
and therefore the decomposition seems beneficial for our purpose
to clarify strengths or ranges of potentials.
At the same time, one finds that
{\it the short-range repulsion arises only from} $V_{\dqex}(R,m_q)$.
Pauli-blocking effect among quarks is essential
for the short-range repulsion.

\section{Discussions}
\label{SecDiscussions}

Now that we have extracted the ingredients of each potential.
Namely, $V_{12,12}(R,m_q)$ and $V_{12,34}(R,m_q)$ are expressed as
\begin{eqnarray}
&&V_{12,34}(R,m_q)=V_{\rm dir}(R,m_q),\\
&&V_{12,12}(R,m_q)=V_{\rm dir}(R,m_q)+V_{\dqex}(R,m_q).
\end{eqnarray}
$V_{\rm dir}(R,m_q)$ and $V_{\dqex}(R,m_q)$
can be regarded as the origins of attractive and repulsive
interactions, respectively.

The key property of the $R$-dependent potentials is twofold;
{\it interaction ranges} and {\it strengths}.
Their $m_q$-dependences are also helpful to clarify
the origins of the potentials.
In phenomenological models,
hadronic interactions are usually incorporated
as one-boson-exchange-potentials (OBEP),
which are Yukawa-type and 
whose interaction ranges are directly connected to (exchanged) meson masses.
Then, such OBE potentials should be sensitive to meson masses,
{\it i.e.} quark masses.
On the other hand, the short-range repulsive core
is often considered to arise from a color-magnetic interaction
induced by one-gluon-exchanges (OGE),
whose strength is proportional to 
the inverse square of (probably constituent) quark masses, $\sim m_{\rm Q}^{-2}$.
Such $m_{\rm Q}$-dependent interactions could be accessed
by monitoring the strengths of the potentials.

In this section,
we evaluate strengths and ranges of these decomposed potentials,
$V_{\rm dir}(R,m_q)$ and $V_{\dqex}(R,m_q)$,
together with their $m_q$-dependences.
Since we do not have priori function forms,
especially for the short-range repulsive part,
we try the following five trial functions,
$F_i(x)$ $(i=1,2,3,4,5)$,
which have two parameters, strength $A$ and range $B$.

\begin{equation}
F_1(x)\equiv A\frac{\exp(-\left(\frac{x}{B}\right)^2)}{\sqrt{\frac{x}{B}}}
\end{equation}
\begin{equation}
F_2(x)\equiv A\frac{\exp(-\left(\frac{x}{B}\right)^2)}{\frac{x}{B}}
\end{equation}
\begin{equation}
F_3(x)\equiv A{\exp\left(-\frac{x}{B}\right)}
\end{equation}
\begin{equation}
F_4(x)\equiv A\frac{\exp(-\frac{x}{B})}{\sqrt{\frac{x}{B}}}
\end{equation}
\begin{equation}
F_5(x)\equiv A\frac{\exp(-\frac{x}{B})}{\frac{x}{B}}
\end{equation}

\begin{figure}[h]
\begin{center}
\includegraphics[scale=0.28]{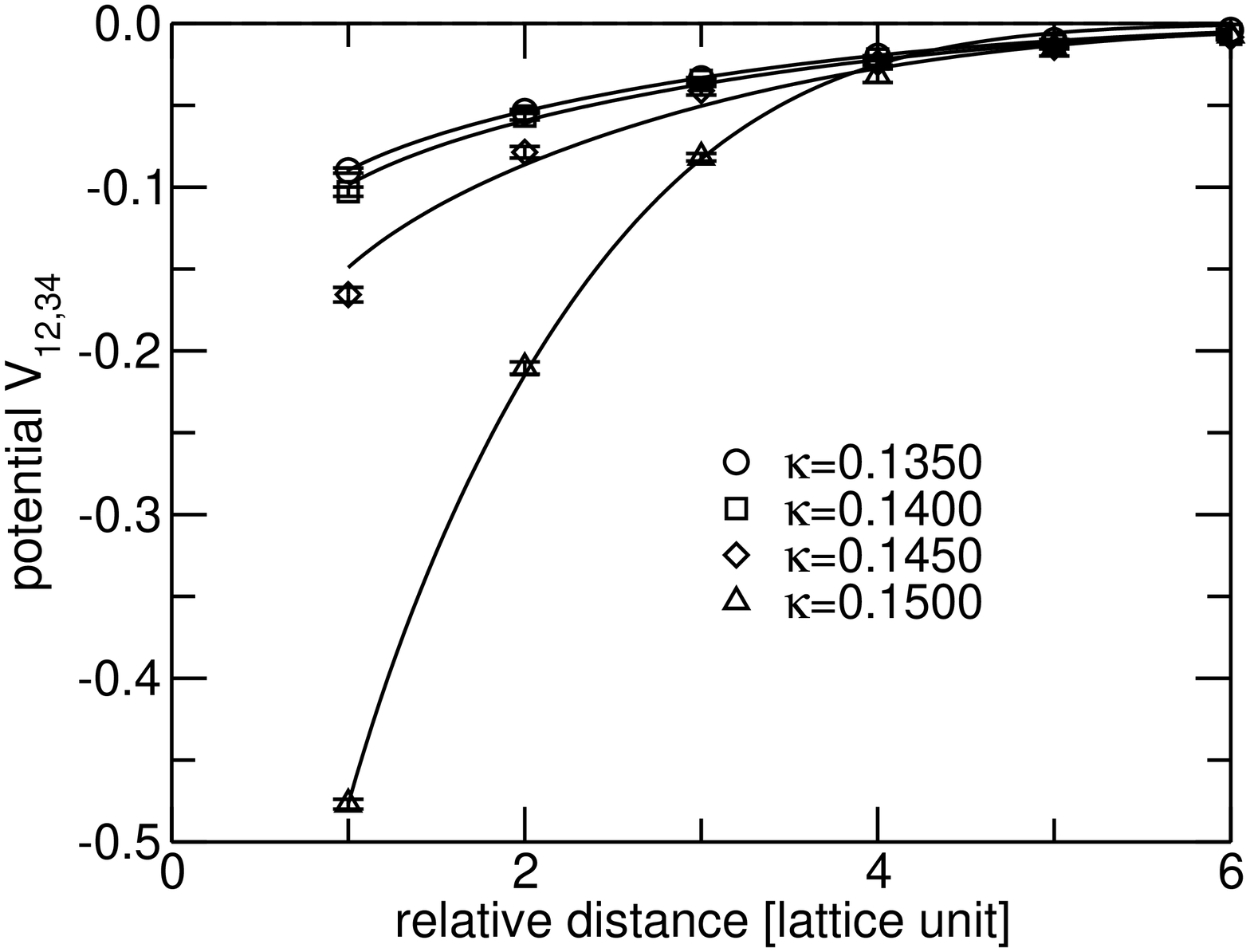}
\includegraphics[scale=0.28]{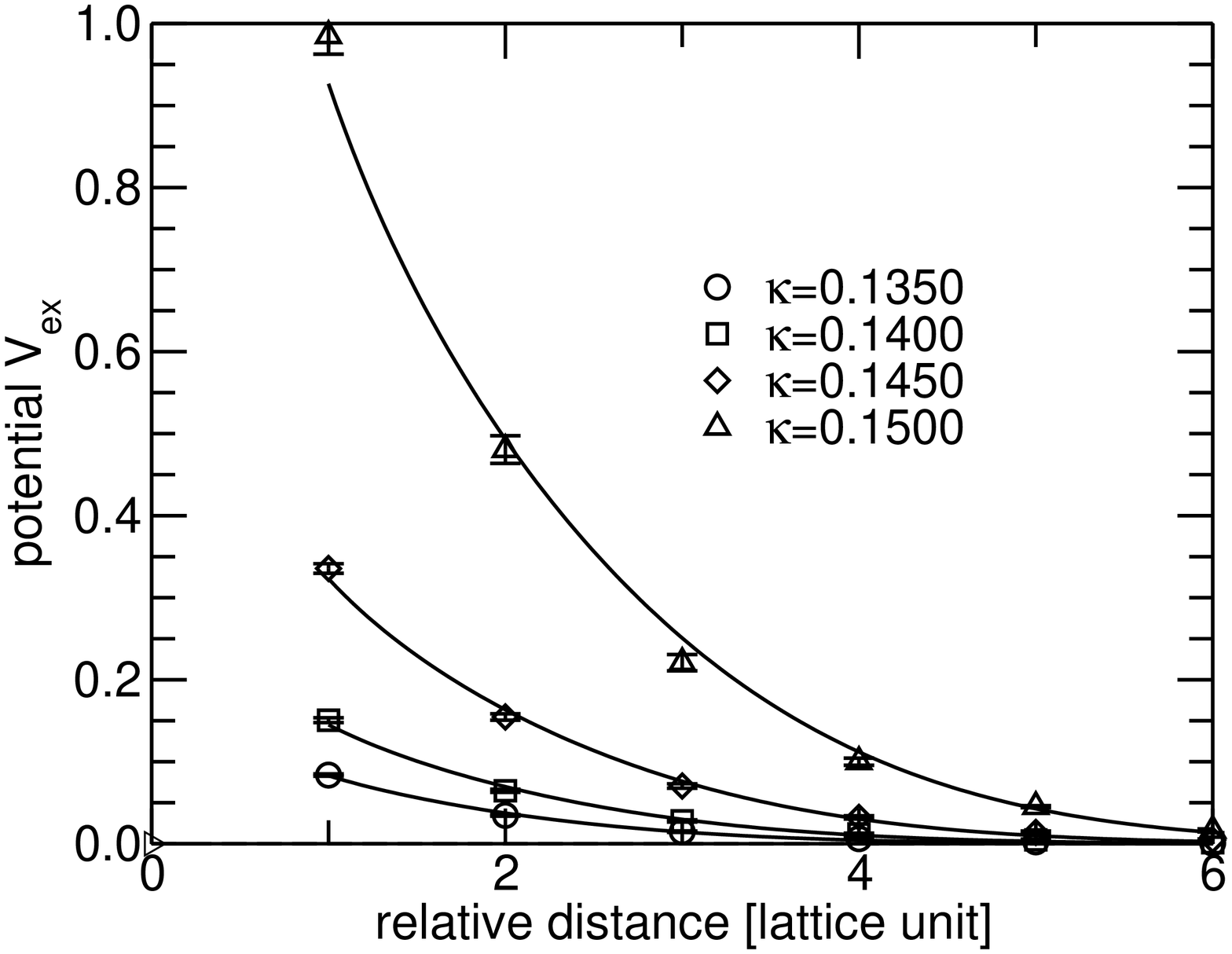}
\end{center}
\caption{
\label{Fighowgood}
{\it Upper}:
Attractive parts of potentials, $V_{\rm dir}(R,m_q)=V_{12,34}(R,m_q)$,
together with the best-fit curves $F_3(R)$
are plotted as functions of relative distance $R$.
{\it Lower}:
``Quark-exchange parts'' of potentials,
$V_{\dqex}(R,m_q)=V_{12,12}(R,m_q)-V_{12,34}(R,m_q)$,
together with the best-fit curves
are plotted as functions of relative distance $R$.
}
\end{figure}

The fit results can be all found in Table.~\ref{fittedparameters},
in which $\chi^2/N_{\rm DF}$, strengths $A$ and ranges $B$
for $V_{\rm dir}(R,m_q)$ and $V_{\dqex}(R,m_q)$ at each $\kappa$ are listed.
None of five functions produces $\chi^2/N_{\rm DF}\simeq 1$
for all the data,
which would be due to the lack of knowledge about detailed function forms,
some systematic errors in the lattice data,
and possible multi-range components, which will be discussed later.
We adopt $F_3$ as the fit function, since it yields
relatively smaller $\chi^2/N_{\rm DF}$ for any combination of potential type
and quark mass.
For reference, we plot (best-fit) $F_3(R)$'s and the lattice data
in Fig.~\ref{Fighowgood}.
The lattice data are rather well mimicked by the function.

\begin{table}[h]
\begin{tabular}{lrrrr}
\hline\hline
$V_{\rm dir}$ & 0.1350 & 0.1400 & 0.1450 & 0.1500 \\ \hline
$\chi^2/N_{\rm df}(F_1)$ & 1.462   & 1.719  & 12.24 & 3.163 \\
$\chi^2/N_{\rm df}(F_2)$ & 54.50   & 6.970  & 0.381 & 9.037 \\
$\chi^2/N_{\rm df}(F_3)$ & 8.938   & 1.094  & 3.516 & 2.206 \\
$\chi^2/N_{\rm df}(F_4)$ & 33.78   & 3.725  & 0.094 & 11.33 \\ 
$\chi^2/N_{\rm df}(F_5)$ & 85.32   & 15.00  & 3.818 & 30.43 \\ \hline

$A(F_1)$ & 0.046(001) & 0.050(002) & 0.081(009) & 0.345(007) \\
$A(F_2)$ & 0.015(004) & 0.017(003) & 0.032(001) & 0.144(010) \\
$A(F_3)$ & 0.154(010) & 0.171(006) & 0.294(019) & 1.134(024) \\
$A(F_4)$ & 0.073(013) & 0.083(008) & 0.162(002) & 0.655(050) \\
$A(F_5)$ & 0.011(008) & 0.015(007) & 0.046(007) & 0.206(052) \\ \hline

$B(F_1)$ & 4.24(006)  & 4.33(011)  & 3.87(024)  & 2.58(005) \\
$B(F_2)$ & 7.06(150)  & 6.77(076)  & 5.36(010)  & 3.61(020) \\
$B(F_3)$ & 1.93(008)  & 1.91(005)  & 1.62(007)  & 1.16(002) \\
$B(F_4)$ & 3.27(041)  & 3.11(021)  & 2.42(002)  & 1.72(011) \\
$B(F_5)$ & 10.3(060)  & 8.12(277)  & 4.63(050)  & 3.20(058) \\ \hline\hline

$V_{\dqex}$ & 0.1350 & 0.1400 & 0.1450 & 0.1500 \\ \hline
$\chi^2/N_{\rm df}(F_1)$ & 3.869  & 5.317  & 6.064 & 7.543 \\
$\chi^2/N_{\rm df}(F_2)$ & 0.546  & 1.144  & 3.713 & 4.562 \\
$\chi^2/N_{\rm df}(F_3)$ & 0.095  & 0.481  & 0.737 & 2.202 \\ 
$\chi^2/N_{\rm df}(F_4)$ & 1.871  & 4.896  & 8.627 & 13.19 \\
$\chi^2/N_{\rm df}(F_5)$ & 8.460  & 16.28  & 27.94 & 33.92 \\ \hline

$A(F_1)$ & 0.060(003)  & 0.099(006)  & 0.209(012) & 0.565(036) \\
$A(F_2)$ & 0.025(001)  & 0.046(002)  & 0.094(006) & 0.277(016) \\
$A(F_3)$ & 0.201(002)  & 0.354(008)  & 0.744(017) & 2.188(085) \\
$A(F_4)$ & 0.119(007)  & 0.223(020)  & 0.453(046) & 1.424(163) \\
$A(F_5)$ & 0.040(010)  & 0.089(023)  & 0.168(050) & 0.606(152) \\ \hline

$B(F_1)$ & 2.58(010)  & 2.78(010)  & 2.99(010) & 3.25(007) \\
$B(F_2)$ & 3.58(008)  & 3.63(009)  & 3.95(015) & 3.97(009) \\
$B(F_3)$ & 1.14(001)  & 1.17(002)  & 1.27(002) & 1.29(002) \\
$B(F_4)$ & 1.67(008)  & 1.64(009)  & 1.80(012) & 1.72(008) \\
$B(F_5)$ & 2.98(048)  & 2.61(041)  & 2.98(055) & 2.60(030) \\ \hline\hline

$V^D_{\rm att}$ & 0.1350 & 0.1400 & 0.1450 & 0.1500 \\ \hline
$\chi^2/N_{\rm df}(F_3)$ & N/A & N/A & 1.338 & 0.246 \\ \hline

$A(F_3)$ & N/A         & N/A         & 0.186(024) & 1.041(046) \\ \hline

$B(F_3)$ & N/A        & N/A        & 1.05(009) & 1.01(004) \\ \hline
\end{tabular}
\caption{\label{fittedparameters}
The best-fit parameters for $V_{\dqex}(R,m_q)$, $V_{\rm dir}(R,m_q)$,
and $V^D_{\rm att}(R,m_q)$. $A(F_i)$ and $B(F_i)$ denote
the strength and the range estimated with the function $F_i$.
}
\end{table}

\subsection{strengths and ranges : $V_{\rm dir}$}

In Table~\ref{fittedparameters} and Fig.~\ref{Figstrrange},
we show the fitted parameters, strength $A_{\rm dir}$ and range $B_{\rm dir}$
of the attractive part $V_{\rm dir}(R,m_q)$,
as functions of half of axialvector-diquark mass.
(Shown as ``DIR'' in Fig.~\ref{Figstrrange}.)
Interestingly enough, both of the strength and the range of
$V_{\rm dir}(R,m_q)$ exhibit flattening in the heavy quark-mass region,
which indicates that
$V_{\rm dir}(R,m_q)$ contains a universal attractive potential
$V^U_{\rm att}(R)$.
``Universal'' here means
that neither the strength nor the range of $V^U_{\rm att}(R)$ depends 
on quark mass
and that $V^U_{\rm att}(R)$ always appears in any flavor channels.
In fact, an attractive ``dip'' at the intermediate range in $V_{12,12}(R,m_q)$ 
seems to originate from this universal attractive potential
$V^U_{\rm att}(R)$.
The combination of $m_q$-dependent short-range repulsive force
and $m_q$-independent long-range attractive force provides a dip
at the intermediate range.

In much heavier quark-mass region,
the properties of quark wavefunctions would undergo gradual change,
and $V^U_{\rm att}(R)$ could be quark-mass dependent.
However, in the quark-mass range we consider here,
$V^U_{\rm att}(R)$ can be treated as a fixed function.

Then, the next question is what the reminder is.
We define the quark-mass dependent part
$V^{D}_{\rm att}(R,m_q)$ as
\begin{equation}
V^{D}_{\rm att}(R,m_q)\equiv
V_{\rm dir}(R,m_q) - V^U_{\rm att}(R).
\end{equation}
We also fit $V^{D}_{\rm att}(R,m_q)$ with $F_3$,
and the fitted parameters are shown
in Table~\ref{fittedparameters} and Fig.~\ref{Figstrrange}.
Here, we simply adopt
$V_{\rm dir}(R,m_q)$ at $\kappa=0.1350$ as the universal part $V^U_{\rm att}(R)$,
since we observe almost no quark-mass dependence already at this $\kappa$.
The strength and the range of $V^{D}_{\rm att}(R,m_q)$
are shown as ``ATT(D)''.
The extracted range $B^D_{\rm att}$ is no longer dependent on quark mass.
That is,
$V_{\rm dir}(R,m_q)$ is approximately described by two independent parts;
\begin{eqnarray}
V_{\rm dir}(R,m_q)
&\sim&
V^U_{\rm att}(R)+V^D_{\rm att}(R,m_q) \\
&=&
A^U_{\rm att} f^U_{\rm att}(R)+
A^D_{\rm att}(m_q) f^D_{\rm att}(R).
\end{eqnarray}

First part represents a $m_q$-independent weak long-range force,
and the second one does a short-range repulsive interaction
that has an $m_q$-dependent strength and an $m_q$-insensitive interaction range.
Possible origins of the universal attraction would be transition
processes to other two hadronic (intermediate) states,
which is schematically illustrated in the left panel of Fig.~\ref{Figexchange},
or $m_q$-independent gluonic interactions.
As the candidates for $m_q$-independent gluonic interactions,
one can consider color-Coulomb interaction,
medium-range attractions due to flavor exchange processes,
and so on, since no quark-mass dependence is observed.
As is found also in our analyses,
such a weak universal attractive force 
is readily masked by the strong repulsive force
in the $V_{12,12}$ potential for lighter quark-mass.

\subsection{strengths and ranges : $V_{\dqex}$}

\begin{figure}[h]
\begin{center}
\includegraphics[scale=0.3]{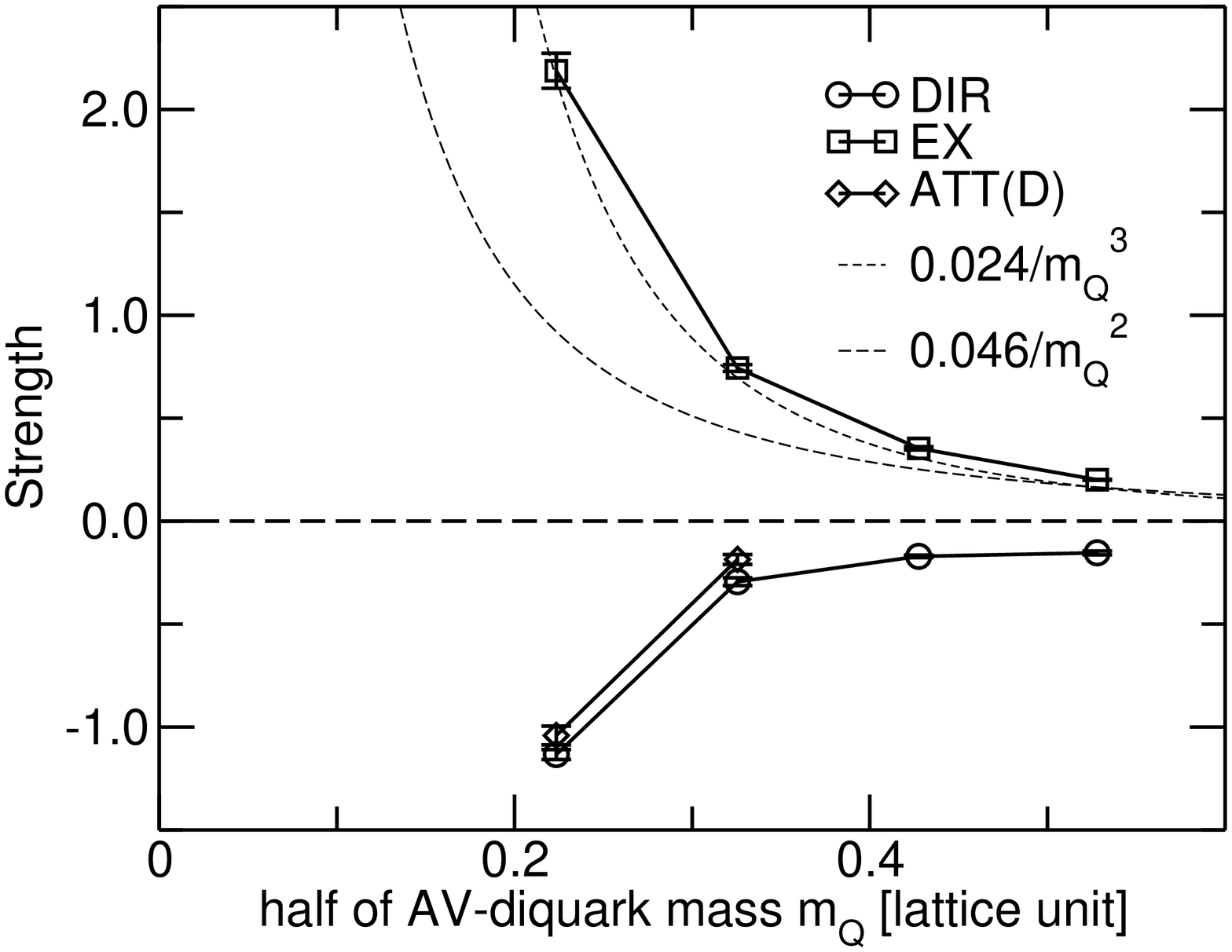}
\includegraphics[scale=0.3]{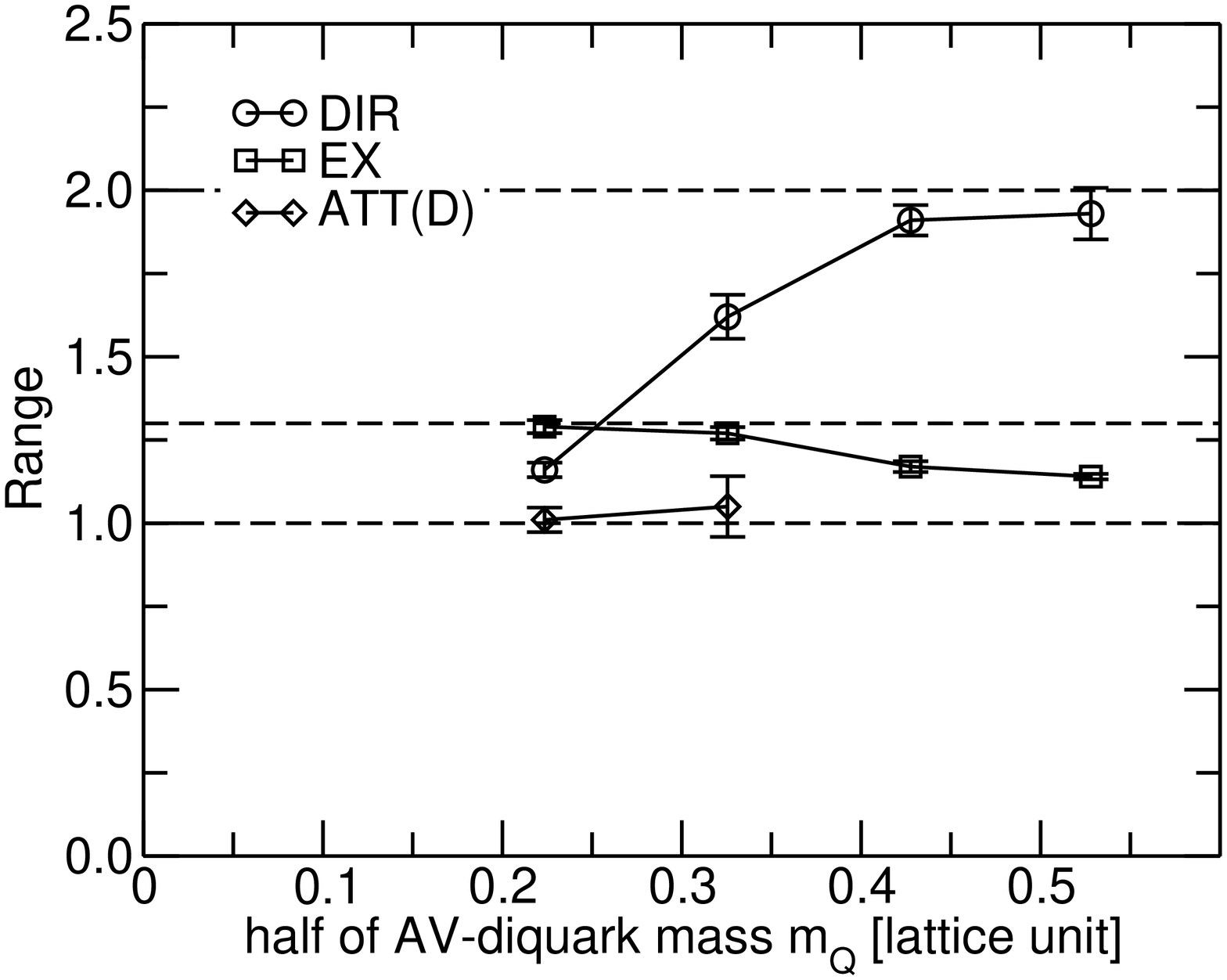}
\end{center}
\caption{
\label{Figstrrange}
{\it Upper}:
The fitted strengths, $A_{\rm dir}$, $A_{\rm ex}$, and $A^D_{\rm att}$,
of the potentials,
$V_{\rm dir}(R,m_q)$, $V_{\dqex}(R,m_q)$, and $V^D_{\rm att}(R,m_q)$,
are plotted as functions of half of axialvector-diquark mass.
{\it Lower}:
The fitted interaction ranges, $B_{\rm dir}$, $B_{\rm ex}$, and $B^D_{\rm att}$,
of the potentials,
$V_{\rm dir}(R,m_q)$, $V_{\dqex}(R,m_q)$, and $V^D_{\rm att}(R,m_q)$.
The parameters for 
$V_{\rm dir}(R,m_q)$, $V_{\dqex}(R,m_q)$, and $V^D_{\rm att}(R,m_q)$
are respectively shown as ``DIR'', ``EX'', and ``ATT(D)''.
}
\end{figure}

In Table~\ref{fittedparameters} and Fig.~\ref{Figstrrange},
we show the fitted parameters, strength $A_{\rm ex}$ and range $B_{\rm ex}$
of the quark-exchange part $V_{\dqex}(R,m_q)$,
as functions of half of axialvector-diquark mass.
(Shown as ``EX'' in Fig.~\ref{Figstrrange}.)
Let us first have a look at the range parameter $B_{\rm ex}$,
which are shown as ``EX'' in Fig.~\ref{Figstrrange}.
One can find that 
the range $B_{\rm ex}$ in $V_{\dqex}(R,m_q)$ is almost quark-mass independent.
It is interesting since
the repulsive cores are sometimes described by {\it e.g.} omega-meson exchanges
in phenomenological models.
The constituent quark-mass 
(half of AV-diquark mass) variation from 0.53 to 0.22
gives rise to only 10\% change in the range $B_{\rm ex}$,
which is much smaller than expected from the meson-mass change.
Moreover, at two largest $\kappa$'s (two lightest quark masses),
the range $B_{\rm ex}$ remains unchanged. 
$V_{\dqex}(R,m_q)$ can be predominantly expressed as
\begin{equation}
V_{\dqex}(R,m_q)
\sim
A_{\rm ex}(m_q)f_{\rm ex}(R).
\end{equation}
The quark-mass dependence (approximately) appears only in the strength $A_{\rm ex}$.
From these observations, we conjecture that the short-range repulsion
is not generated by meson poles
as OBEP, but by some other origins.

One most possible mechanism for the short-range repulsion
is a color-magnetic (CM) interaction among quarks
as suggested by constituent-quark models.
A CM interaction accompanied by Pauli blocking among quarks 
raises the energy of (closely located) two hadrons,
which results in strong repulsions.
In the first-order perturbation,
the contributions from the CM interaction are proportional to 
its strength, whose $m_q$-dependence is given by $m_{\rm Q}^{-2}$.

The strengths $A_{\rm ex}$ extracted by fits are also shown
in Table~\ref{fittedparameters} and Fig.~\ref{Figstrrange}.
$A_{\rm ex}$ increases in the light-quark-mass region
and it is qualitatively consistent with the CM interaction as the origin of repulsion.
In the upper panel in Fig.~\ref{Figstrrange}, two fit functions,
$Cm_{\rm Q}^{-2}$ and $Cm_{\rm Q}^{-3}$ are respectively plotted as
a long-dashed and a dashed line.
The strength $A_{\rm ex}$ is well reproduced by
$Cm_{\rm Q}^{-3}$ rather than $Cm_{\rm Q}^{-2}$,
which implies that the quark-mass dependence of repulsion
seems stronger than that of the strength of the CM interaction itself.
However, considering that the Pauli blocking among quarks is essential 
for the repulsion and that the interaction range of the repulsive force has a
very weak quark-mass dependence,
the present results are consistent with
the quark-model interpretation that short-range repulsion between hadrons
arises from CM interaction among quarks.

\section{Some additional trials}
\label{SecSome}

\subsection{Boundary condition dependences}

\begin{figure}[h]
\begin{center}
\includegraphics[scale=0.45]{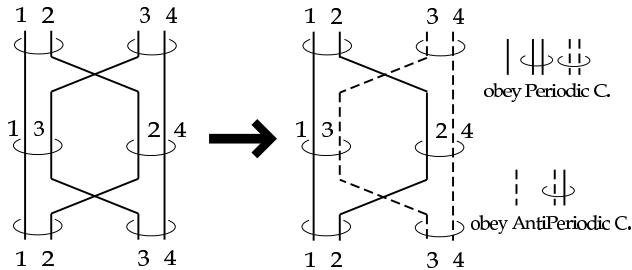}
\end{center}
\caption{
\label{Figexchange}
Schematic figure for quark-exchange interactions.
Two hadrons, whose flavors are initially and finally $(1,2)$ and $(3,4)$,
interact with each other through the intermediate state,
where their flavors are $(1,3)$ and $(2,4)$, for example.
If we impose the antiperiodic boundary condition
onto 3- and 4-quarks,
$(1,2)$- and $(3,4)$-diquarks obey the periodic boundary condition,
whereas the intermediate $(1,3)$- and $(2,4)$-diquarks
do the antiperiodic boundary condition.
As a result, the energies of the intermediate states are raised up,
if they are spatially spread and being affected by finite-size effects.
}
\end{figure}

In the previous section,
we claimed that one possible contribution for 
the universal attraction would be transition
processes to other two hadronic (intermediate) states,
which is schematically illustrated in the left panel of Fig.~\ref{Figexchange}.
In case that two hadrons are closely located,
it may not be meaningful to identify their flavors,
because in such short-distance regions,
two hadrons are largely overlapping with each other
and they may differ from two isolated hadrons~\cite{Koike:1986mm,Okiharu:2004ve}.

In order to get sure that we have no finite-volume artifact
for the universal attraction,
we repeat the same analyses 
with antiperiodic spatial boundary conditions~\cite{Ishii:2004qe,Takahashi:2009bu}.
The flavors are again $(i,j,k,l)=(1,2,3,4)$,
and anti-periodic boundary conditions are imposed
for $3,4$-flavor quark fields.
In this case, $(1,2)$-$(3,4)$ two-diquark state is never affected,
since $(1,2)$- and $(3,4)$-diquarks obey a periodic boundary condition.
If there exists finite-volume artifact in intermediate transition states,
{\it e.g.} $(1,3)$-$(2,4)$ hadronic state,
the results would be modified in this new analysis.
We eventually observed no modification in all the cases,
and no serious finite volume effect is confirmed.

\subsection{Removal of high-momentum gluons}
\label{SecRemoval}

Hadronic interactions may be classified into two categories.
One is highly nonperturbative phenomena, such as meson interactions,
and the other is perturbative contributions represented as
CM interactions among quarks.
In one sense, these phenomena differs in energy scale they belong to.
Recently, a method that restricts energy scale of gluons
via the Fourier transformation was proposed and tested~\cite{Yamamoto:2008am,Yamamoto:2008ze}.
The authors reported that
removal of high-energy contributions results in
vanishing short-range Coulomb interaction in heavy-quark potentials,
keeping the linear confinement part still unchanged.
Taking into account that the Coulomb potential in short-distance region
comes from OGE processes, removal of high-energy gluons
is expected to cut such OGE processes.
Though it should be clarified whether such removal also cuts
other OGE interactions, {\it e.g.} color-magnetic part,
this analyses might give us
a hint to short-range hadronic interactions.
Especially, the origin of short-range repulsive core
coming from the color-magnetic OGE interaction
would be accessed to some extent in this manner.

We perform 3-dim Fourier transformation~\cite{Yamamoto:2009na} 
for link variables $U_\mu ({\bf x},t)$ in the Landau gauge
and extract $U_\mu ({\bf p},t)$ in momentum space:
\begin{equation}
U_\mu ({\bf p},t)
=
\sum_{\bf x}
U_\mu ({\bf x},t)
e^{i{\bf p}\cdot{\bf x}}
\end{equation}
We leave the low-momentum contribution,
\begin{equation}
\widetilde{U}^{\rm low}_\mu ({\bf p},t)=
\begin{cases}
\; U_\mu ({\bf p},t)\ (|{\bf p}|\leq\Lambda) \\
\; 0\ ({\rm otherwise})
\end{cases}
\end{equation}
and reconstruct new SU(2) link variables in the coordinate space,
$U^{\rm low}_\mu ({\bf x},t)$, so that the distance,
\begin{equation}
{\rm Tr} \ 
(U^{\rm low}_\mu ({\bf x},t)-\widetilde{U}^{\rm low}_\mu ({\bf x},t))
(U^{\rm low}_\mu ({\bf x},t)-\widetilde{U}^{\rm low}_\mu ({\bf x},t))^\dagger,
\end{equation}
is minimized.
Here, 
\begin{equation}
\widetilde{U}^{\rm low}_\mu ({\bf x},t)\equiv
\frac{1}{V}\sum_{\bf p}\widetilde{U}^{\rm low}_\mu ({\bf p},t)e^{-i{\bf p}\cdot{\bf x}}
\end{equation}
is a truncated link variable reconverted into the coordinate representation,
which no longer belongs to SU(2).
The infrared cut $\Lambda$ 
is set to $\Lambda = 5$ in lattice unit.

\begin{figure}[h]
\begin{center}
\includegraphics[scale=0.3]{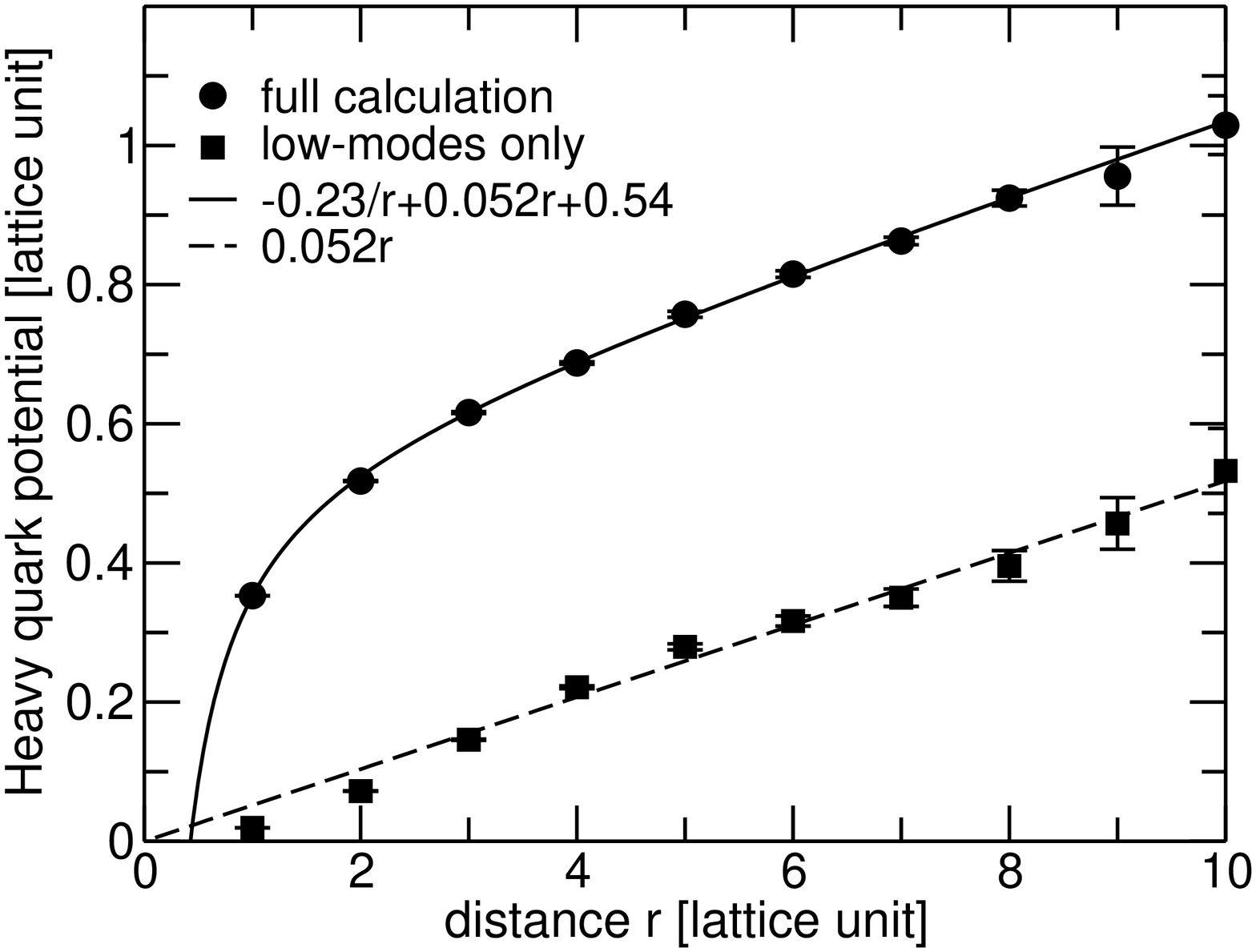}
\end{center}
\caption{
\label{Figwloop}
Heavy quark-antiquark potential as a function of a separation $r$.
The upper data shown as circles denote the lattice QCD data
obtained in full SU(2) calculation, and
the lower data shown as squares denote those
obtained without high-momentum gluons.
The infrared cut $\Lambda$ is set to $\Lambda = 5$ in lattice unit.
The original $Q\bar Q$ potential is fitted
as $V_{Q\bar Q}(r)=-0.2397(53)\frac1r + 0.0518(10)r + 0.5405(50)$.
}
\end{figure}

We show the interquark potential in Fig.~\ref{Figwloop}.
The upper data in Fig.~\ref{Figwloop} 
shown as circles denote the lattice QCD data $V_{Q\bar Q}(r)$
obtained in full SU(2) calculation, and
the lower data shown as squares denote those $V^{\rm low}_{Q\bar Q}(r)$
obtained without high-momentum gluons.
The original $Q\bar Q$-potential data $V_{Q\bar Q}(r)$ is fitted
with the Cornell type potential as,
$V_{Q\bar Q}(r)=-0.2397(53)\frac1r + 0.0518(10)r + 0.5405(50)$.
On the other hand, a fit of $V^{\rm low}_{Q\bar Q}(r)$
using the Cornell type potential
seems not applicable due to the oscillation 
behavior in the short-range regions.
We simply draw a dotted line, $V(r)=0.0518r$, in the figure.
One can find that the short-distance Coulomb interaction 
disappears and the linear confinement part remains almost unchanged.
At the same time, the constant part in the potential,
which represents the self energy of a static fundamental charge 
under the lattice regularization,
also disappears in this treatment.

An interesting observation can be found in the masses
of scalar and axialvector diquarks.
We adopt two hopping parameters $\kappa$ = 0.1150 and 0.1250.
We note here that we have adjusted $\kappa$'s
to cover approximately the same quark-mass region
so that the AV-diquark masses before and after high-momentum-gluon cut
are similar.
In Table.~\ref{hadronicmasses}, we list the masses
of scalar(S) and axialvector(AV) diquarks.
The masses measured with low-momentum link variables
almost show no mass splitting,
which implies that the color-magnetic interaction
that gives rise to S-AV mass splitting is now largely suppressed.
The fact that S- and AV-diquark masses degenerate
in this manner again implies the validity of
nonrelativistic-quark-model description of the diquarks.
The NG-boson nature in S-diquark hardly appears
in the present quark-mass region.

\begin{figure}[htb]
\begin{center}
\includegraphics[scale=0.3]{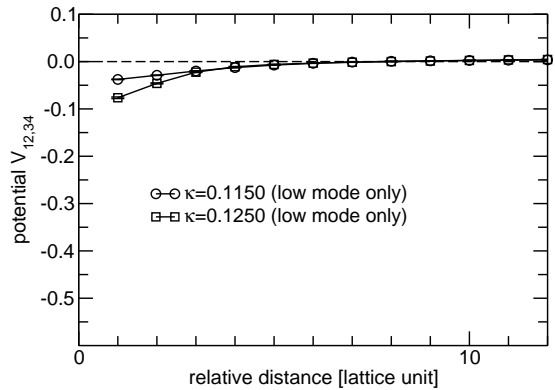}
\end{center}
\caption{
\label{Figpotentials_diag1.woge}
Potentials $V_{12,34}(R)$ for the flavor combination $(i,j,k,l)=(1,2,3,4)$
computed without high-momentum gluons
are plotted as functions of relative distance $R$.
The infrared cut $\Lambda$ is set to $\Lambda = 5$ in lattice unit.
}
\end{figure}
\begin{figure}
\includegraphics[scale=0.3]{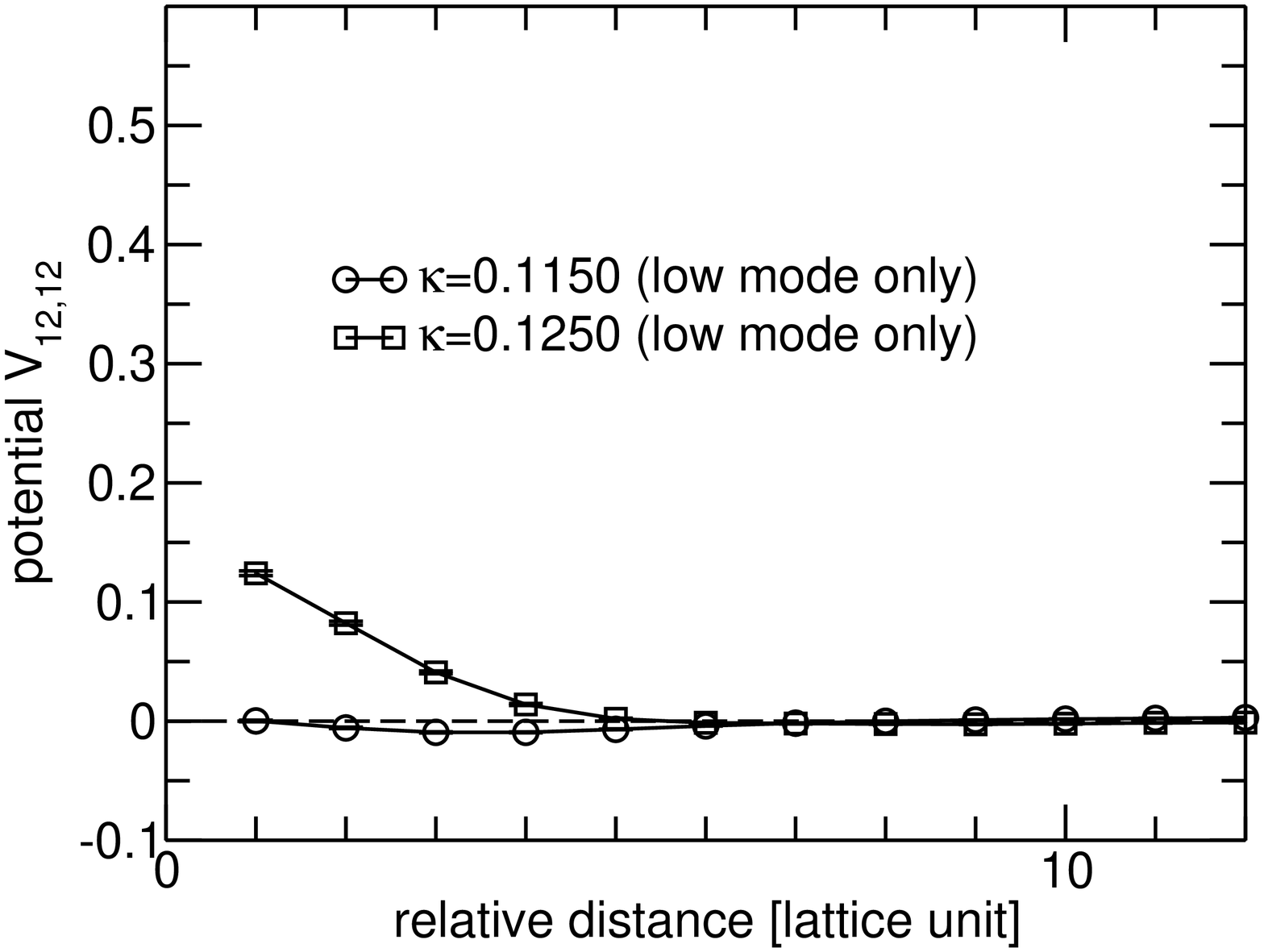}
\caption{
\label{Figpotentials_diag12.woge}
Potentials $V_{12,12}(R)$ for the flavor combination $(i,j,k,l)=(1,2,1,2)$
computed without high-momentum gluons
are plotted as functions of relative distance $R$.
The infrared cut $\Lambda$ is set to $\Lambda = 5$ in lattice unit.
}
\end{figure}
\begin{figure}
\begin{center}
\includegraphics[scale=0.3]{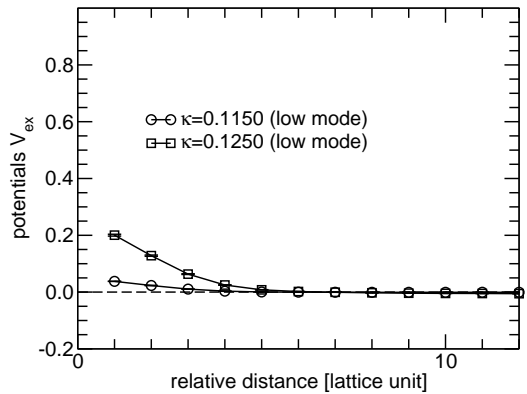}
\end{center}
\caption{
\label{Figpotentials_diff.woge}
``Quark-exchange parts'' of potentials, which are defined as 
$V_{\dqex}(R)\equiv V_{12,12}(R)-V_{12,34}(R)$,
computed without high-momentum gluons
are plotted as functions of relative distance $R$.
The infrared cut $\Lambda$ is set to $\Lambda = 5$ in lattice unit.
}
\end{figure}

We show the interhadron potentials 
measured with the low-momentum link variables in 
Figs.~\ref{Figpotentials_diag1.woge},~\ref{Figpotentials_diag12.woge},
and ~\ref{Figpotentials_diff.woge}.
Naive expectation is that
potentials originating from short-range OGE processes are suppressed,
while those from other origins remain unchanged.
Actually, all the potentials get smaller 
after the high-momentum gluons are removed,
which can be seen in 
Figs.~\ref{Figpotentials_diag1.woge},~\ref{Figpotentials_diag12.woge},
and ~\ref{Figpotentials_diff.woge}.
The $R$-dependence of potentials are also much milder now,
and the divergent behavior around the origin disappears.
The reason why all are now decreased could be that
many of the hadronic processes,
{\it e.g.} flavor exchanges, meson exchanges, and so on,
need color recombinations in hadrons.
If such recombinations are prohibited,
some of hadronic processes depending on them 
will be also suppressed.

\begin{figure}[htb]
\begin{center}
\includegraphics[scale=0.7]{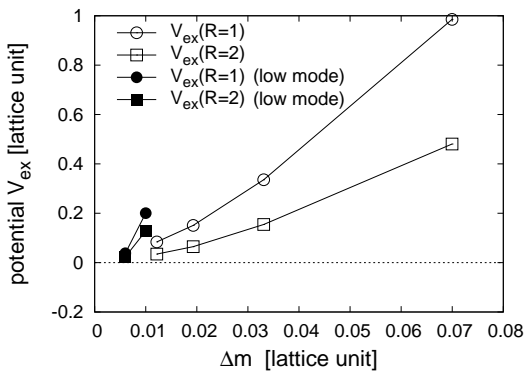}
\end{center}
\caption{
\label{Figpot-ex}
The values of $V_{\dqex}$ at $R$=1 and 2
are plotted as functions of S-AV mass splitting $\Delta m$.
}
\end{figure}

To clarify the origin of repulsion in a clearer way,
we plot the values of $V_{\dqex}$ at $R$=1 and 2
as functions of S-AV mass splitting $\Delta m$
in Fig.~\ref{Figpot-ex}.
The open symbols are the original data
taken from fully calculated potential $V_{\dqex}$,
and the filled symbols denote the values of $V_{\dqex}$
obtained without high-momentum gluons.
Open and filled symbols are compared at the same $\Delta m$.
While the open symbols show $\Delta m^{\frac32}\simeq m_{\rm Q}^{-3}$ behavior,
the filled symbols deviate from the expected curve
and always lie above it, which indicates that
the repulsive interaction does {\it not always} decrease
like the S-AV mass splitting,
and that some extra enhancement remains
even after the high-momentum-gluon cut.
We then expect that the repulsive force still contains
small contributions other than CM interaction among quarks.

Restricting gluon's energy scale surely changes potentials
and may be useful to clarify detailed structures.
However, there seem to remain several issues to be clarified
before quantitative conclusions are reached.

\subsection{Different choice in operators}

Concern about possible operator dependences
of potentials from BS amplitudes
has been often raised~\cite{Beane:2010em}.
BS amplitudes themselves are operator-dependent quantities,
and inevitably equivalent potentials extracted from them are.
Though the strengths and the ranges 
of the potentials would be operator dependent,
some essential features in hadronic interactions
will be clarified by investigating operator dependences.
Among possible hadronic operators,
we hereby consider spatially extended smeared operators,
because hadron-size effect is considered 
to be one of the most important features:
Hadronic states have finite sizes, and hence finite-size (smeared)
hadronic operators would be more adequate to describe hadrons.
We note that the reduction formula needed to relate
BS amplitudes to scattering observables has been proved
only for point-type operators~\cite{Aoki:2009ji,Nishijima:1958zz,Zimmermann:1958hg,Haag:1958vt}.
In this subsection, we repeat our analyses employing 
smeared operators for sinks
in order to derive essentials in hadronic interactions.

First, we mention smearing methods.
Usually, quark operators are individually smeared,
and hadronic operators are constructed from them.
In the case of gauge-invariant Gaussian smearing,
each quark operator is smeared with a gauge-covariant
lattice derivative operator 
\begin{equation}
K_\mu(U)_{xy}
\equiv
U_\mu^\dagger(x)
\delta_{x+\hat \mu,y}
+
U_\mu(x-\hat \mu)
\delta_{x-\hat \mu,y}
\end{equation}
as
\begin{eqnarray}
q_b({\bf x};{\bf r})
&\equiv&
\prod^{N}
\left(
1+\alpha \sum_{i=1}^3 K_i(U)_{\bf xy}
\right)
\delta_{{\bf y},{\bf r}}.
\end{eqnarray}
Parameters $\alpha$ and $N$ are chosen so that
quark fields are distributed around ${\bf r}$ with a radius $b$.
We hereby consider diquark states which consist of 
two quarks with identical masses.
A smeared hadronic (diquark) operator $H_b({\bf x_1},{\bf x_2};{\bf x})$ 
whose center position is $x$ is then defined
with $q_{1, b}({\bf x_1};{\bf x})$ and $q_{2, b}({\bf x_2};{\bf x})$
as
$H_b({\bf x_1},{\bf x_2};{\bf x})=q_{1, b}({\bf x_1};{\bf x})q_{2,
b}({\bf x_2};{\bf x})$.
However, in reality, $H_b({\bf x_1},{\bf x_2};{\bf x})$ does not describe
a hadronic state located at ${\bf x}$,
but it contains unwanted contributions
because a ``position'' of a hadron, $({\bf x_1}+{\bf x_2})/2$, 
does not always coincide with ${\bf x}$.
For example, $H_b({\bf x_1},{\bf x_2};{\bf x})$ has nonzero entry at
${\bf x_1}={\bf x_2}={\bf x'}\neq {\bf x}$,
which is nothing but a point-type hadronic operator
located at ${\bf x'}\neq {\bf x}$.
While such contributions are not harmful in hadron-mass measurement
because hadronic states are projected onto a momentum eigenstate
summing up hadronic correlators,
it causes serious difference in the measurement of BS amplitudes.
$H_b({\bf x_1},{\bf x_2};{\bf x})$ should be zero when 
$({\bf x_1}+{\bf x_2})/2\neq {\bf x}$.
In order to satisfy the condition $({\bf x_1}+{\bf x_2})/2= {\bf x}$,
we improve smearing function with the following prescription.

With new link variables,
$U^{\theta_\mu}_\mu\equiv \exp(i\theta_\mu)U_\mu$,
and a phase $0 \leq \theta_\mu < 2\pi$,
we define
\begin{eqnarray}
&&K^{\theta_\mu}_\mu(U)_{xy} \\
&\equiv&U^{\theta_\mu\dagger}_\mu(x)
\delta_{x+\hat \mu,y}
+
U^{\theta_\mu}_\mu(x-\hat \mu)
\delta_{x-\hat \mu,y} \\
&=&\exp(-i{\theta_\mu})U_\mu^\dagger(x)
\delta_{x+\hat \mu,y}
+
\exp(i{\theta_\mu})U_\mu(x-\hat \mu)
\delta_{x-\hat \mu,y}.
\end{eqnarray}
Smeared quark operators are similarly defined as
\begin{eqnarray}
q^{\vec \theta}_b({\bf x};{\bf r})
&\equiv&
\prod^{N}
\left(
1+\alpha \sum_{i=1}^3 K^{\theta_i}_i(U)_{\bf xy}
\right)
\delta_{{\bf y},{\bf r}}.
\end{eqnarray}
Here, $\vec \theta = (\theta_1,\theta_2,\theta_3)$
can be chosen freely.
As a result,
$q^{\vec \theta}_b({\bf x};{\bf r})$ are different
from $q_b({\bf x};{\bf r})$ in overall phases;
\begin{equation}
q^{\vec \theta}_b({\bf x};{\bf r})
=
\exp(i\vec \theta \cdot ({\bf x-r}))q_b({\bf x};{\bf r}).
\end{equation}
For simplicity, we hereby consider a hadronic operator located at the origin, 
${\bf x}={\bf 0}$.
Then, a hadronic (diquark) operator
$H^{\vec \theta_1 \vec \theta_2}_b({\bf x_1},{\bf x_2};{\bf 0})$ is now 
\begin{eqnarray}
&&H^{\vec \theta_1 \vec \theta_2}_b({\bf x_1},{\bf x_2};{\bf 0}) \\
&=&
q^{\vec \theta_1}_{1, b}({\bf x_1};{\bf 0})q^{\vec \theta_2}_{2,b}({\bf
x_2};{\bf 0}) \\
&=&
\exp(i\vec \theta_1 \cdot {\bf x_1}+i\vec \theta_2 \cdot {\bf x_2})
q_{1, b}({\bf x_1};{\bf 0})q_{2,b}({\bf x_2};{\bf 0})
.
\end{eqnarray}
Taking $\vec \theta_1=\vec \theta_2=\vec \theta$,
\begin{eqnarray}
&&H^{\vec \theta}_b({\bf x_1},{\bf x_2};{\bf 0}) \\
&=&
\exp(i\vec \theta \cdot ({\bf x_1}+{\bf x_2}))
q_{1, b}({\bf x_1};{\bf 0})q_{2,b}({\bf x_2};{\bf 0})
.
\end{eqnarray}
If we average 
$H^{\vec \theta}_b({\bf x_1},{\bf x_2};{\bf 0})$
as $\sum_{\vec \theta} H^{\vec \theta}_b({\bf x_1},{\bf x_2};{\bf 0})$
over $\vec \theta$ so that
$\exp(i\vec \theta \cdot ({\bf x_1}+{\bf x_2}))=\delta_{{\bf x_1}+{\bf
x_2},{\bf 0}}$,
the ``position'' of a hadronic operator can be unambiguously defined.
(${\bf x_1}+{\bf x_2}={\bf 0}$ is always satisfied.)
The simplest choice is averaging it over a random set $\{\vec \theta\}$.
This scheme is a generalization of projection processes, 
and could be employed for other purposes.

We note here that nothing other is changed in this process.
A hadronic operator is now ``projected'' onto
(${\bf x_1}+{\bf x_2}={\bf 0}$)-state
so that unwanted contributions are absent.
In actual calculations,
$H^{\vec \theta}_b({\bf x_1},{\bf x_2};{\bf 0})$
are averaged over the set 
$\{\vec \theta \}=\{\frac{2\pi}{N_\theta}(l,m,n)\} (1\leq l,m,n \leq N_\theta)$.
When $N_\theta$ is smaller than the lattice spatial extent,
the delta function $\delta_{{\bf x_1}+{\bf x_2},{\bf 0}}$
is not fully reproduced and the projection is incomplete.
To eliminate such (small) contaminations,
we further introduce overall random phases.

\begin{figure}[h]
\begin{center}
\includegraphics[scale=0.3]{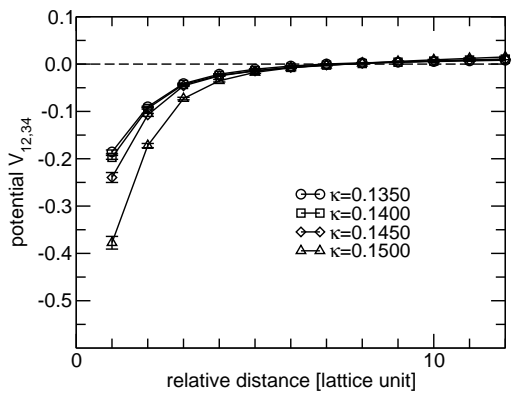}
\end{center}
\caption{
\label{Figpotentials_diag1_cmsmr30}
Potentials $V_{12,34}(R)$ computed 
with ``projected'' smeared diquark operators
$\sum_{\vec \theta}H^{\vec \theta}_b$,
with the flavor combination, $(i,j,k,l)=(1,2,3,4)$,
are plotted as functions of relative distance $R$.}
\end{figure}

\begin{figure}[h]
\begin{center}
\includegraphics[scale=0.3]{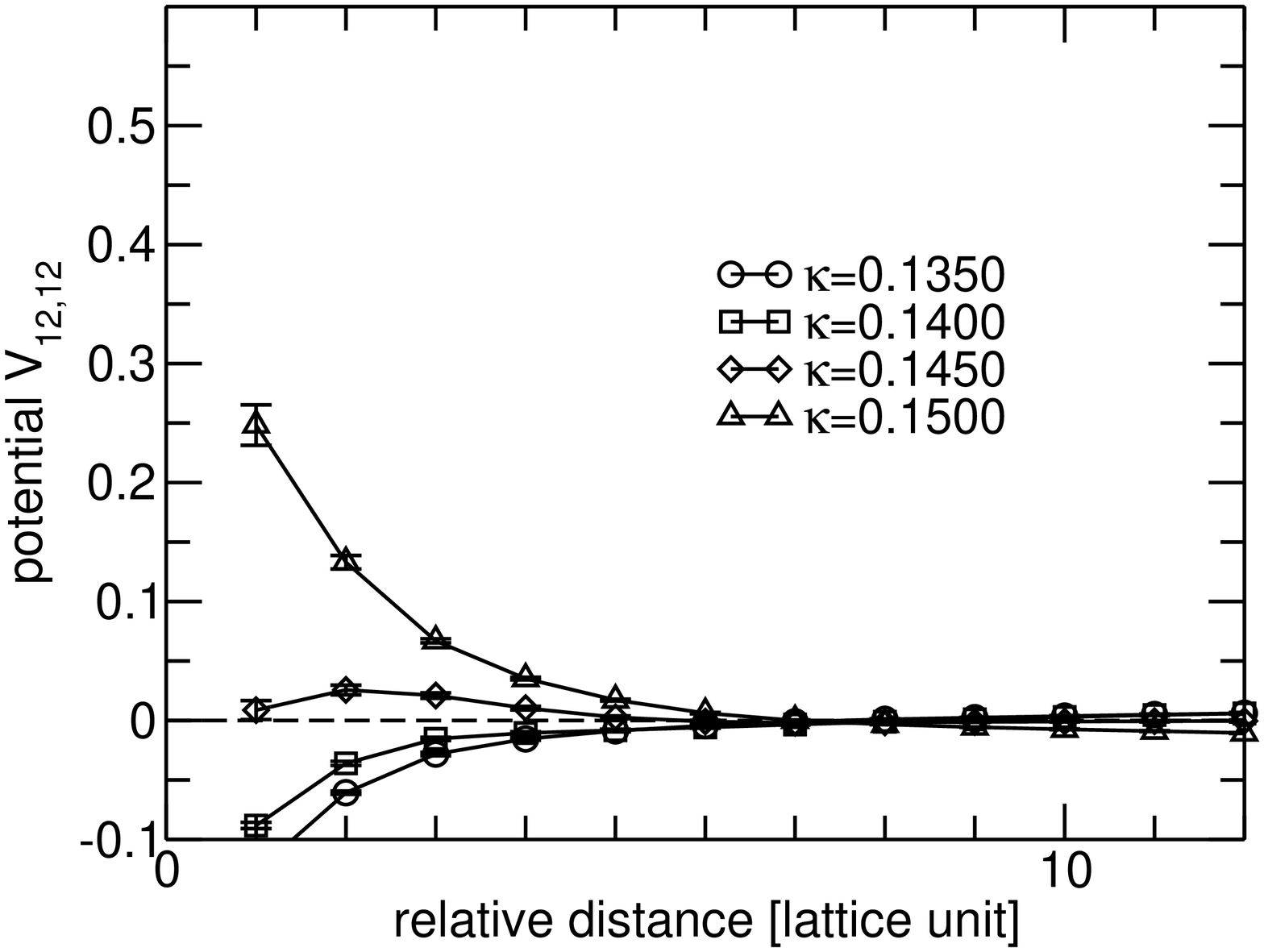}
\end{center}
\caption{
Potentials $V_{12,12}(R)$ computed 
with ``projected'' smeared diquark operators
$\sum_{\vec \theta}H^{\vec \theta}_b$,
with the flavor combination, $(i,j,k,l)=(1,2,1,2)$,
are plotted as functions of relative distance $R$.
\label{Figpotentials_diag12_cmsmr30}}
\end{figure}

\begin{figure}[h]
\begin{center}
\includegraphics[scale=0.3]{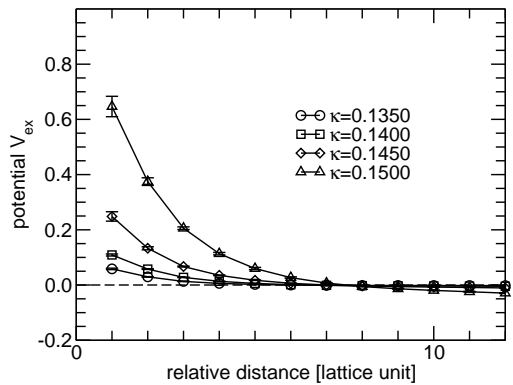}
\end{center}
\caption{
\label{Figpotentials_diff_cmsmr30}
``Quark-exchange parts'' of potentials,
which are defined as 
$V_{\dqex}(R)\equiv V_{12,12}(R)-V_{12,34}(R)$,
are plotted as functions of relative distance $R$.
In this case, $V_{12,12}(R)$ and $V_{12,34}(R)$
are those measured 
with ``projected'' smeared diquark operators
$\sum_{\vec \theta}H^{\vec \theta}_b$.
}
\end{figure}

Next, we measure hadron ``sizes'' monitoring overlap coefficients.
The sink operators relevant to BS amplitudes
are now smeared in a gauge invariant way.
The root-mean-square radius $b_{\rm smr}$ of the smeared operators
are determined so that the smeared operators have maximal overlaps
with the ground states (scalar diquarks).
\begin{figure}[h]
\begin{center}
\includegraphics[scale=0.28]{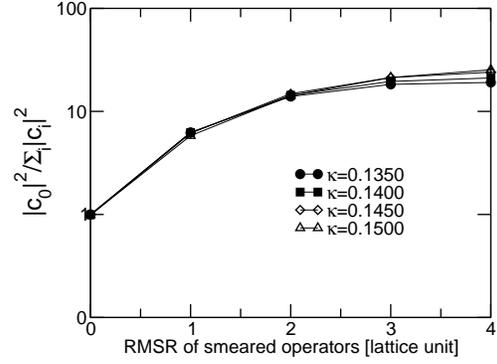}
\end{center}
\caption{
$|c_0|^2/\sum_i |c_i|^2$ are plotted as a function of 
the RMSR of a smeared operator $b_{\rm smr}$.
Here, $c_i$ denotes the overlap
of a diquark operator and $i$-th state;
$c_i\equiv \langle{\rm vac}|D_{ij,\Gamma}|i{\rm th \  state}\rangle$.
\label{Figoverlap}}
\end{figure}
In Fig.~\ref{Figoverlap}, we plot 
the squared overlaps, $|c_0|^2/\sum_i |c_i|^2$.
Here, $c_i$ denotes the overlap
of a diquark operator and $i$-th state, which is defined as
\begin{equation}
c_i\equiv \langle{\rm vac}|
D_{ij,\Gamma}
|i{\rm th \  state}\rangle.
\end{equation}
While the squared overlaps are slightly dependent on quark masses,
they are saturated at $b_{\rm smr}\sim 3$.
We then set $b_{\rm smr}= 3$ for all the $\kappa$'s.
In Figs.~\ref{Figpotentials_diag1_cmsmr30}-\ref{Figpotentials_diff_cmsmr30},
we show $V_{12,34}$,$V_{12,12}$, and $V_{\rm ex}$
obtained with ``properly projected'' smeared operators.

$V_{12,34}$ obtained with smeared operators
are qualitatively similar to those with point operators.
It is always attractive and gets stronger at smaller quark-mass region.
Especially, the saturation at heavy $m_Q$ region can be again seen
in Fig.~\ref{Figpotentials_diag1_cmsmr30}.
The interaction range of the universal attractive potential
$V_{\rm att}^U(R)$ is now reduced by about 20\%,
and the strength is stronger, compared with the universal attraction
observed in the point-operator case.
On the other hand,
repulsive cores in $V_{12,12}$ at short-distance region
are generally weaker in this case, which is displayed in
Fig.~\ref{Figpotentials_diag12_cmsmr30}.
Apparent repulsive cores disappear at small $\kappa$'s,
whereas cores still persist and are observed at larger $\kappa$'s.
$V_{12,12}$ seems more sensitive to operator choice than $V_{12,34}$.
Interestingly, the difference 
$V_{\dqex}(R)\equiv V_{12,12}(R)-V_{12,34}(R)$
still hold similar properties to that in point-operator case.
It grows as $m_Q$ decreases and is always repulsive,
which confirms that the Pauli-blocking effect 
is responsible for repulsion also in this case.
The interaction range of $V_{\dqex}$ remains almost unchanged
as compared to that extracted with point operators.

Quantitative evaluation of interhadron potentials
seems difficult at present, because of its operator dependences.
However, several basic properties
remain qualitatively unchanged with/without operator smearing:
(1) {\it The existence of universal attraction between two hadrons.}
(2) {\it The difference $V_{\dqex}(R)\equiv V_{12,12}(R)-V_{12,34}(R)$
still shows repulsive contribution at any $\kappa$'s.}
(3) {\it The quark-mass dependences of the interaction ranges of
direct ($V_{12,34}$) and exchange ($V_{\dqex}$) parts of potentials 
are much weaker than expected from meson exchanges.}
(4) {\it The quark-mass dependence of the strength of $V_{\dqex}$ is 
consistent with or stronger than $m_Q^{-2}$.}

\begin{figure}[h]
\begin{center}
\includegraphics[scale=0.3]{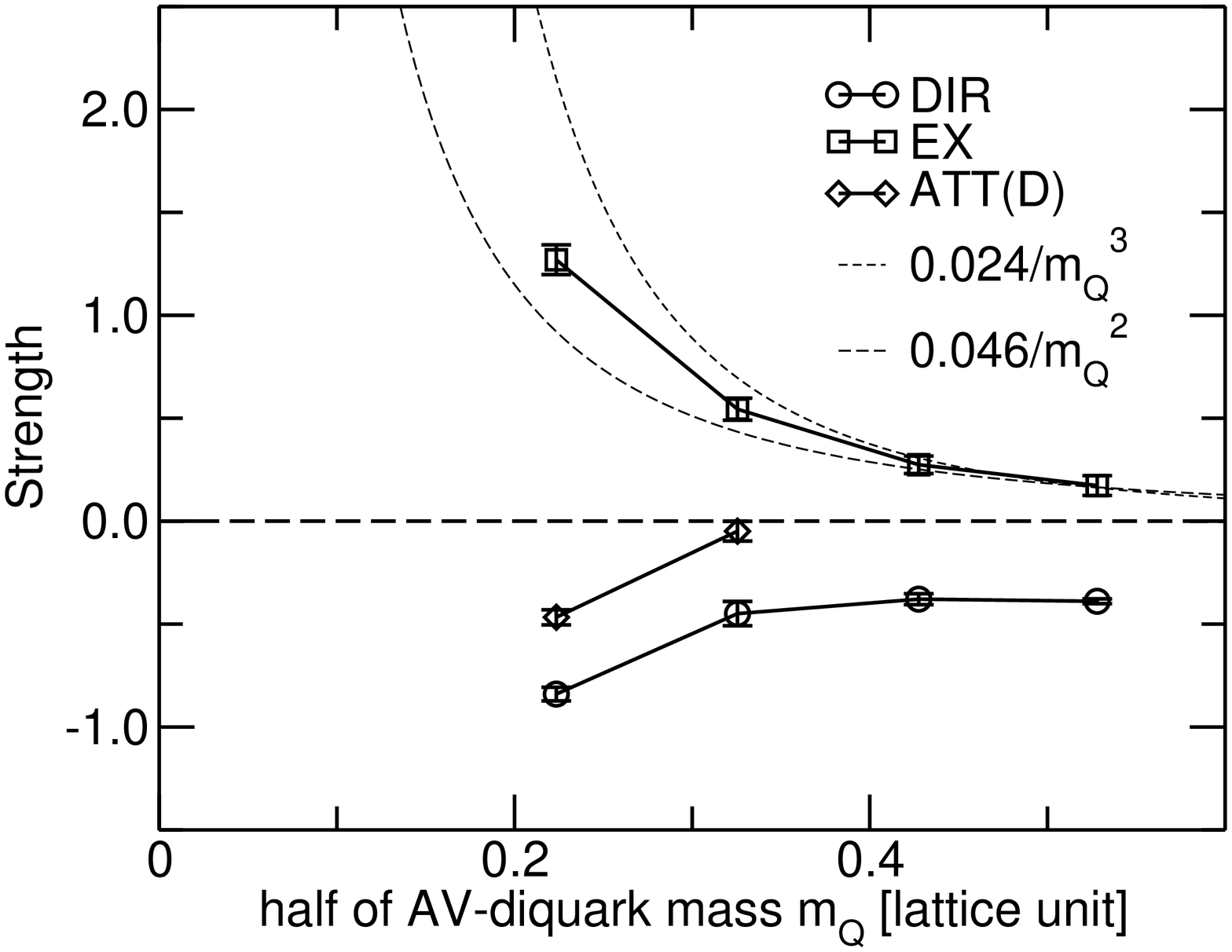}
\includegraphics[scale=0.3]{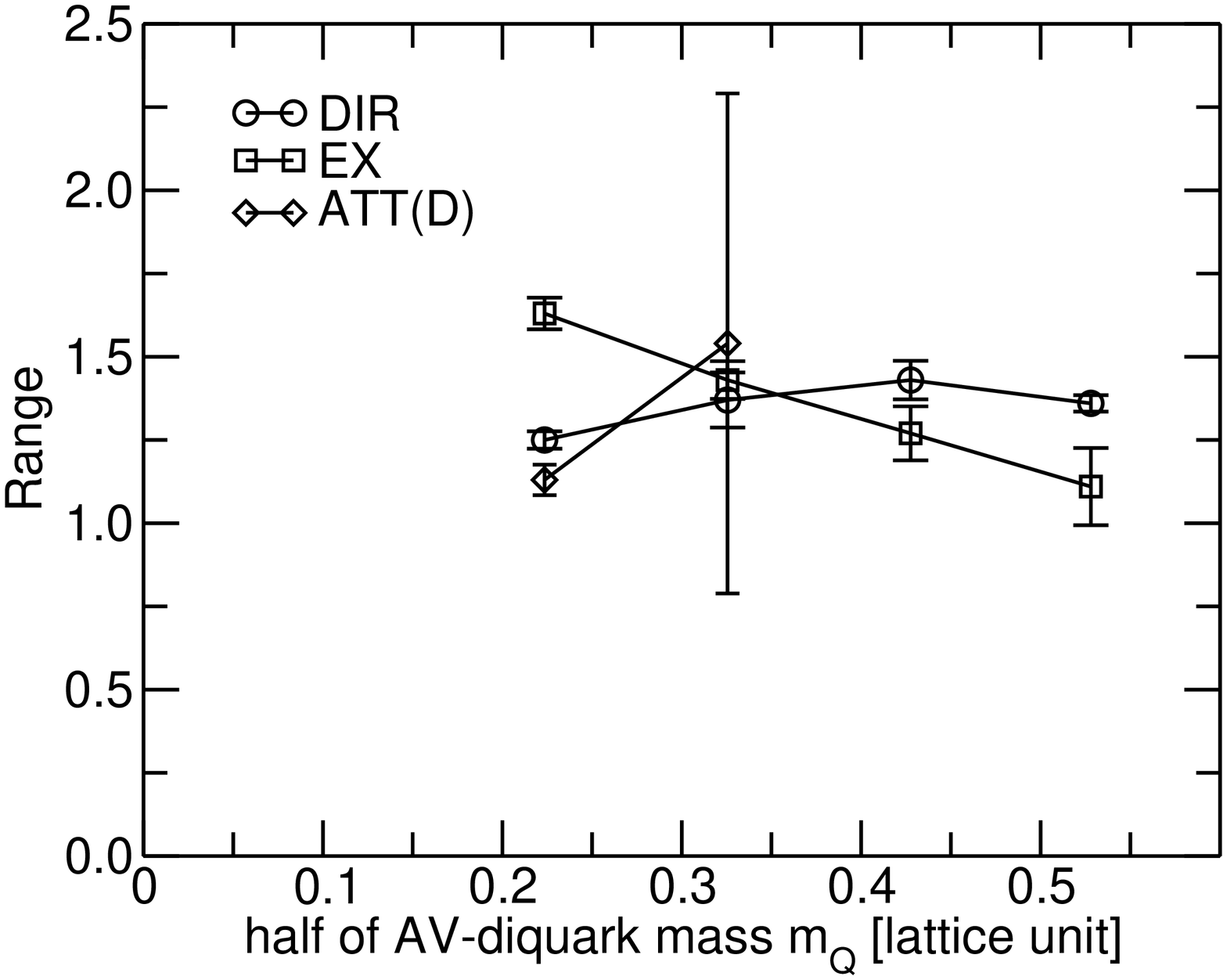}
\end{center}
\caption{
\label{Figstrrange_cmsmr30}
{\it Upper}:
The fitted strengths, $A_{\rm dir}$, $A_{\rm ex}$, and $A^D_{\rm att}$,
of the potentials,
$V_{\rm dir}(R,m_q)$, $V_{\dqex}(R,m_q)$, and $V^D_{\rm att}(R,m_q)$,
are plotted as functions of half of axialvector-diquark mass.
{\it Lower}:
The fitted interaction ranges, $B_{\rm dir}$, $B_{\rm ex}$, and $B^D_{\rm att}$,
of the potentials,
$V_{\rm dir}(R,m_q)$, $V_{\dqex}(R,m_q)$, and $V^D_{\rm att}(R,m_q)$.
The parameters for 
$V_{\rm dir}(R,m_q)$, $V_{\dqex}(R,m_q)$, and $V^D_{\rm att}(R,m_q)$
are respectively shown as ``DIR'', ``EX'', and ``ATT(D)''. \\
All the parameters are obtained by fitting potentials
measured
with ``projected'' smeared diquark operators
$\sum_{\vec \theta}H^{\vec \theta}_b$.
}
\end{figure}

\section{Summary and Outlooks}
\label{SecSummary}

We have evaluated inter-hadron interactions
in two-color lattice QCD
based on the $R$-dependent function $V(R)$.
The function $V(R)$,
which has been expressed as ``potential'' throughout this paper,
has been extracted from
the Bethe-Salpeter wave functions on the lattice
assuming a nonrelativistic Schr{\" o}dinger equation.
The simulations have been performed in quenched two-color QCD
with the plaquette gauge action at $\beta = 2.45$ and the Wilson quark action.
Evaluating different flavor combinations as well as several quark masses,
we have extracted and investigated each ingredient in hadronic interactions.
We have considered two-diquark scatterings,
whose flavors are $(i,j)$ and $(k,l)$.
When all the quarks have different flavors,
$(i,j,k,l)=(1,2,3,4)$,
the interhadron potential is quark-mass dependent and always attractive,
\begin{equation}
V_{12,34}(R,m_q) = V_{\rm dir}(R,m_q).
\end{equation}
On the other hand, in $(i,j,k,l)=(1,2,1,2)$ channel,
where exchange diagrams, {\it i.e.} Pauli-blocking effect
among quarks, are included,
$V_{12,12}(R,m_q)$ shows repulsion in short-distance region,
and this repulsion arises only from the exchange diagrams.
We have defined the ``quark-exchange part'' $V_{\dqex}(R,m_q)$
in the potential,
which is induced by adding quark-exchange diagrams,
or equivalently, by introducing Pauli blocking among some of quarks.
Namely, when some of the quarks have identical flavors,
$(i,j,k,l)=(1,2,1,2)$,
the quark-exchange part $V_{\dqex}(R,m_q)$
is defined as $V_{\dqex}(R,m_q)\equiv V_{12,12}(R,m_q)-V_{12,34}(R,m_q)$,
which means the interhadron potential is written as
\begin{equation}
V_{12,12}(R,m_q) = V_{\rm dir}(R,m_q) + V_{\dqex}(R,m_q).
\end{equation}
It indicates that the short-range repulsive core
in $V_{12,12}(R,m_q)$
{\it arises only from the ``quark-exchange part''} $V_{\dqex}(R,m_q)$.
Pauli blocking among quarks seems essentially needed
to generate repulsive force between two hadrons.

We have found that these 
$R$- and $m_q$-dependent contributions can be further decomposed
into simpler parts;
\begin{eqnarray}
V_{\dqex}(R,m_q)
&\sim&
A_{\rm ex}(m_q)f_{\rm ex}(R), \\
V_{\rm dir}(R,m_q)
&\sim&
V^U_{\rm att}(R)+V^D_{\rm att}(R,m_q) \\
&=&
A^U_{\rm att} f^U_{\rm att}(R)+
A^D_{\rm att}(m_q) f^D_{\rm att}(R).
\end{eqnarray}
Here, $f$'s are quark-mass independent functions of $R$.
Interestingly, $V_{\rm dir}(R,m_q)$ can be further decomposed into
$V^U_{\rm att}(R)$ and $V^D_{\rm att}(R,m_q)$,
a universal attractive part and an $m_q$-dependent attractive part.

One prominent observation is that
all the $R$-functions $f_{\rm ex}(R)$, 
$f^U_{\rm att}(R)$, and $f^D_{\rm att}(R)$
are quark-mass independent,
which implies that {\it the quark-mass dependences of}
$V_{\dqex}(R,m_q)$ and $V^D_{\rm att}(R,m_q)$
{\it do not appear in interaction ranges but only in the overall strengths}.
Though we first adopted the specific function $F_3(x)$
for potential-form analyses, we finally found that
simple rescaling of the potential $V_{\dqex}(R,m_a)$
(the lattice data themselves)
well explains all the quark-mass dependences.
That is, we eventually encounter no scheme dependence
in these descriptions.
These observations apparently 
do not go together with the meson-exchange picture among hadrons,
since $m_q$-dependences of interaction ranges seem very small.
At least in the present quark-mass range,
mesonic contributions seem small and subdominant,
and hence we observed interactions of the other origins.

For the repulsive part,
we have found that the strength grows as $\sim m_{\rm Q}^{-3}$,
which is similar to but slightly stronger than
the color-magnetic interaction itself by one-gluon-exchange (OGE) processes.
Considering that the Pauli blocking among quarks is essential 
for the repulsion and 
the interaction range of the repulsive force has a
very weak quark-mass dependence,
it is likely to originate from a color-magnetic interaction among quarks.

Actually, the function $V(R)$ is energy and operator dependent.
In order to clarify operator dependence,
we have constructed ``projected'' smeared operators
and compared the results with and without operator smearing.
As a result, we observed quantitative changes in potential shapes.
Quantitative evaluation of interhadron potentials
seems difficult at present, because of its operator dependences.
On the other hand, several basic properties
remain qualitatively unchanged with/without operator smearing:
(1) {\it The existence of universal attraction between two hadrons.}
(2) {\it The difference $V_{\dqex}(R)\equiv V_{12,12}(R)-V_{12,34}(R)$
still shows repulsive contribution at any $\kappa$'s.}
(3) {\it The quark-mass dependences of the interaction ranges of 
direct and exchange parts of potentials ($V_{\rm dir}$,$V_{\dqex}$)
are much weaker than expected from meson exchanges.}
(4) {\it The quark-mass dependence of the strength of $V_{\dqex}$ is
consistent with or stronger than $m_Q^{-2}$.}
To determine the precise forms of interhadron potentials,
it would be necessary to perform more sophisticated analyses
taking into account energy dependence
in a less operator-dependent way.

Since quark-mass dependence in interaction ranges does not appear,
the origin of the interactions we observed so far 
would be all gluonic interactions and/or flavor-exchange processes,
{\it e.g.} color-magnetic or color-Coulomb interactions, and so on,
rather than mesonic contributions.
As was found in our analyses,
attractive forces are readily masked by the strong repulsive force
appearing in the $V_{12,12}$ potential in the light quark-mass region.
If the universal attraction in hadronic interaction appears 
also in SU(3) QCD,
they might be observed in a channel with
no or less quark-exchange contribution
between hadrons, such as $N\phi$ scattering state.
In the lighter quark-mass region,
meson-exchange contributions could largely emerge and be predominant,
which is left for further studies.

\acknowledgments
All the numerical calculations were performed on NEC SX-8R at CMC, Osaka university, on SX-8 at YITP, Kyoto University. The authors thank Dr. K. Yazaki for discussions. This work was supported in part by the Yukawa International Program for Quark-Hadron Sciences (YIPQS), and the Grant-in-Aid for the Global COE 
Program ``The Next Generation of Physics, 
Spun from Universality and Emergence'' from MEXT of Japan,
and by KAKENHI (20028006, 21740181).

\end{document}